\documentclass[useAMS,usenatbib,graphicx]{mn2e}

\usepackage{graphicx}

\title[MHD simulation of clump and filament formation]{MHD simulation of the formation of clumps and filaments 
in quiescent diffuse medium by thermal instability}
\author[C. J. Wareing et al.]{C. J. Wareing$^{1}$\thanks{E-mail:
C.J.Wareing@leeds.ac.uk}, J. M. Pittard$^{1}$, S. A. E. G. Falle$^{2}$ and S. Van Loo$^{1}$\\
$^{1}$School of Physics and Astronomy, University of Leeds, Leeds, LS2 9JT, U.K.\\
$^{2}$School of Mathematics, University of Leeds, Leeds, LS2 9JT, U.K.}
\begin{document}

\date{Accepted 2016 March 08. Received 2016 March 01; in original form 2016 January 15}

\pagerange{\pageref{firstpage}--\pageref{lastpage}} \pubyear{2002}

\maketitle

\label{firstpage}

\begin{abstract}
We have used the AMR hydrodynamic code, MG, to perform idealised 3D MHD simulations of the formation of 
clumpy and filamentary structure in a thermally unstable medium without turbulence. A stationary thermally unstable 
spherical diffuse atomic cloud with uniform density in pressure equilibrium with low density surroundings was 
seeded with random density variations and allowed to evolve. A range of magnetic field strengths 
threading the cloud have been explored, from $\beta=0.1$ to $\beta=1.0$ to the zero magnetic
field case ($\beta=\infty$), where $\beta$ is the ratio of thermal pressure to magnetic
pressure. Once the density inhomogeneities had developed to the point where 
gravity started to become important, self-gravity was introduced to the simulation. With no
magnetic field, clouds and clumps form within the cloud with aspect ratios of around unity, whereas in the 
presence of a relatively strong field ($\beta=0.1$) these become filaments, then evolve into 
interconnected corrugated sheets that are predominantly perpendicular to the magnetic field. 
With magnetic and thermal pressure equality ($\beta=1.0$), filaments, clouds and clumps
are formed. At any particular instant, the projection of the 3D structure onto a plane parallel 
to the magnetic field, i.e. a line of sight {\it perpendicular} to the magnetic field, resembles the 
appearance of filamentary molecular clouds. The filament densities, widths, velocity dispersions 
and temperatures resemble those observed in molecular clouds. 
In contrast, in the strong field case $\beta=0.1$, projection of the 3D structure along a 
line of sight {\it parallel} to the magnetic field reveals a remarkably uniform structure.
\end{abstract}


\section{Introduction}

Extensive studies of the nearest star-forming clouds, most recently with the {\it Herschel} Space
Observatory have revealed that every interstellar cloud contains an intricate network of interconnecting
filamentary structures (see, for example, Section 2 of the review of \cite{andre14} and references 
therein). 
The data, from {\it Herschel} and near-IR studies for example, suggest a scenario 
in which these ubiquitous filaments represent a key step in the star formation process: large-scale flows
compress the diffuse ISM and form molecular clouds; an interconnecting filamentary structure forms
within these clouds; magnetic fields affect the directions of movement and hence overall structure, although 
do not appear to set the central densities in the filaments; gravity plays an increasingly important role, 
fragmenting the filaments once they are cold and dense into prestellar cores and finally protostars.
Observational results now connect well with numerical simulations, as
highlighted in Section 5 of \cite{andre14} and the references therein. Numerical
simulations now include the thermodynamic behaviour of the cloud material, magnetic fields, gravity
and feedback from massive stars, both radiative and dynamic. Turbulence has emerged as an ingredient
which can, injected at the right scale, result in the formation of filaments which possess properties remarkably
similar to those derived from observational results. 

In the work presented here, we explore the formation of filaments through the use of MHD simulations
of the thermal instability \citep{field65} in a low-density cloud of quiescent diffuse 
medium initially in the warm unstable phase and in
pressure equilibrium with its lower-density surroundings, including accurate thermodynamics, magnetic fields 
and self-gravity. Our motivation is to study the underlying physics of the thermal instability under 
initially quiescent conditions, without additional complications such as driven turbulence or colliding flows.
In the next two sections, we review recent relevant results, summarising properties of filaments derived 
from observational results in Section \ref{obsreview}   
and recent relevant work on analytical and numerical filament formation models in Section 
\ref{modreview}. In Section \ref{numerical} we present our numerical method and define the initial 
conditions used in our model. In Section \ref{results} we present our results and in 
Section \ref{analysis} analyse those results with comparison with the observational results discussed in 
Section \ref{obsreview}. We conclude the work in Section \ref{conclusions}.


\section{Properties of filaments}\label{obsreview}

Filaments and their importance for star formation have been noted by many authors for decades. We 
follow the review of \citet{andre14} and define filaments as any elongated ISM structures with an 
aspect ratio larger than $\sim$5-10 that are significantly overdense with respect to their surroundings.
They are not thought to be projections of sheets or larger structures.
\citet{schneider79} discussed the properties of elongated dark nebulae with internal structures they
named ``globular filaments". Within star-forming molecular gas, CO and dust observations revealed
that both the Orion A cloud \citep[e.g.][]{bally87,chini97,johnstone99} and the Taurus Cloud 
\citep[e.g.][]{abergel94,mizuno95,hartmann02,nutter08,goldsmith08} have prominent filamentary structure.
Other well-known examples include the molecular clouds in the constellations Musca and Chamaeleon \citep[e.g.][]{cambresy99},
Perseus \citep[e.g.][]{hatchell05}, and S106 \citep*[e.g.][]{balsara01}. After making comparisons,
\cite{myers09} noted that young stellar groups and clusters are frequently associated with dense ``hubs"
radiating multiple lower-column-density filaments.

The {\it Herschel} Space Observatory has now uncovered filamentary structure in molecular clouds
and infrared dark clouds in great detail (see the review of \citet{andre14} and also, for example, \citet{andre10},
\citet{konyves10}, \citet{arzou11,arzou13}, \citet{peretto12}, \citet{schneider12} and \citet{palm13} for more detail).
Notably, filamentary structure is present in every cloud observed with {\it Herschel}, independent of
whether the cloud is actively star-forming or not. For example, the Polaris Flare, a translucent, 
non-star forming cloud is clearly filamentary in structure in both {\it Herschel}-derived column
density and in 250 $\mu$m continuum emission \citep{ward10,miville10}. This ubiquity indicates
the formation of filaments precedes star formation in the cold ISM, and is tied to processes
acting within clouds themselves \citep{andre14}.

The {\it Herschel} observations have revealed a number of interesting results regarding the properties
of filaments. Most often noted is the result from detailed analysis of resolved filamentary column 
density profiles \citep[e.g.][]{arzou11,juvela12,palm13} that the shape of filament profiles is universal
and described by a Plummer-like function of the form \citep{plummer1911,whitworth01,nutter08,arzou11}:
\begin{equation}
\displaystyle{{\rho _p}\left( r \right) = \frac{{{\rho _c}}}{{{{\left[ {1 + ({r \mathord{\left/
 {\vphantom {r {{R_{flat}}{)^2}}}} \right.
 \kern-\nulldelimiterspace} {{R_{flat}}{)^2}}}} \right]}^{{p \mathord{\left/
 {\vphantom {p 2}} \right.
 \kern-\nulldelimiterspace} 2}}}}}}
\end{equation}
for the density profile, equivalent to:
\begin{equation}
\displaystyle{{\Sigma _p}\left( r \right) = {A_p}\frac{{{\rho _c}{R_{flat}}}}{{{{\left[ {1 + ({r \mathord{\left/
 {\vphantom {r {{R_{flat}}{)^2}}}} \right.
 \kern-\nulldelimiterspace} {{R_{flat}}{)^2}}}} \right]}^{\frac{{p - 1}}{2}}}}}}
\end{equation}
for the column density profile, where $\rho_c$ is the central density of the filament, $R_{flat}$ is the
radius of the flat inner region, $ p \approx 2$ is the power-law exponent at large radii ($r >> R_{flat}$),
and $A_p$ is a finite constant factor which includes the effect of the filament's inclination angle to the
plane of the sky. It is notable that the exponent $p$ is not 4, which would be the case for an 
isothermal gas cylinder in hydrostatic equilibrium \citep{ostriker64}. \cite{palm13} introduced a 
possible explanation for why $p \approx 2$ at large radii, explaining that dense filaments may not
be strictly isothermal, but may be better described by a polytropic equation of state, 
$P \propto \rho^{\gamma}$ or $T \propto \rho^{\gamma-1}$ with $\gamma \leq 1$. 
Observational measurement of the mean dust temperature profile measured perpendicular to the
B213/B211 filament in Taurus shows the best polytropic model fit to temperature is achieved with
a polytropic index $\gamma=0.97\pm0.01$ \citep[see][for further details]{palm13}. Dust temperatures 
in the filament are on the order of 10-15K and this model assumes $T_{gas}=T_{dust}$.
However, filaments may be more dynamic systems than the static equilibrium assumed therein.

\cite{arzou11} also found that when averaged over the length of the filaments, the diameter
$ 2 \times R_{flat}$ of the flat inner plateau in the radial profiles for 27 filaments in Gould 
belt clouds is a remarkably constant $0.1 \pm 0.03$\,pc. Arzoumanian et al. conclude there is no correlation 
between filament width and central column density for the Gould belt clouds. Considering
this further, from their figure 7, filament widths range from the half peak beam width (HPBW)
resolution limit of 0.03\,pc up to 0.1\,pc in Polaris, from the HPBW resolution limit of 0.05\,pc to 
0.2\,pc in Aquila and from the HPBW resolution limit of 0.08\,pc to 0.2\,pc in IC5146.
Other authors have found larger values for filament full widths at half-maximum (FWHM).
\cite{hennemann12} found widths between 0.26\,pc and 0.34\,pc for the DR21 ridge and
filaments in Cygnus X and similarly \cite{juvela12} found FWHM of around 0.32\,pc for
filaments within the {\it Planck} Galactic cold cores.

Line emission observations of C$^{18}$O and N$_{2}$H$^{+}$ and other molecules towards star forming filaments 
\citep*{zuc74,arzou13,hacar13,furuya14,henshaw14,jimenez14,li14} have revealed non-thermal line 
broadening. 
\cite{arzou13} presented molecular line measurements of the internal velocity dispersions in
46 {\it Herschel} identified filaments. Noting the thermal sound speed of $\sim$0.2\,km\,s$^{-1}$
for T=10\,K, they found velocity dispersions in the range 0.2-0.4\,km\,s$^{-1}$ for thermally subcritical 
and nearly critical filaments, implying the level of ``turbulent" motions is almost constant and does
not dominate over thermal support. For thermally supercritical filaments (i.e. they contain more
mass than the thermal pressure can withhold), they find a positive
correlation between filament column density and the level of turbulent motions, observing velocity
dispersions up to 0.6\,km\,s$^{-1}$. This points to an additional source
driving these motions, generally regarded as turbulent, the origin of which is unclear. First interpreted as indications of
gravitational collapse \citep{gold74}, the scales were rapidly noted to be too small, but
several mechanisms have since been proposed as the source, including protostellar outflows,
expanding HII regions, stellar winds and internal supernovae (SNe), external SNe, colliding flows or tidal
forces and accretion and collapse 
\citep[see][and references therein for a more complete discussion]{ibanez15}.
Observations indicate that the driving source is on the largest scales in molecular clouds
\citep*{maclow00,brunt03,brunt09} making it difficult for internal point sources, e.g. stellar feedback, to drive the
large-scale flows. Self-gravity or multiple combined SNe have recently emerged as leading candidates
\citep{ibanez15} but the debate is not settled. 

Large-scale and well-ordered magnetic fields have been revealed through polarisation measurements
towards star-forming filaments. In many cases, the magnetic field appears to be roughly perpendicular
to the filaments \citep[e.g.][]{chapman11,sugitani11,planck14a}. There are some reports though of a
bimodal distribution of field directions - either parallel or perpendicular to the major axis of the
filament itself \citep{li13,pillai15,planck14b}. 
Lower density filaments parallel to the magnetic field have been 
coined ``striations" marking the flow of material along field lines accreting onto the denser 
perpendicular filaments to which they are connected \citep{hacar13}.

Analytical studies have also shed light on the stability and fragmentation properties of filaments
subject to turbulent motions, external pressure confinement and accretion
on the filament \citep*[see, e.g.][]{ostriker64,inutsuka92,fischera12,pon11,pon12,toala12,heitsch13}.
Magnetic fields have generally been found to have a positive effect on filament stability 
\citep{nagasawa87,fiege00,heitsch13,tomisaka14}. \citet{soler13} combined numerical simulations
of magnetised molecular clouds and synthetic polarisation maps in order to show that the relative
orientation of the magnetic field also depends on the initial magnetisation of the filament-forming
cloud. This conclusion has been used to infer details of the driving process: super-Alfv{\'e}nic 
turbulence causes strong compression, resulting in magnetic fields parallel to the filamentary
structures \citep{padoan01}, whereas sub-Alfv{\'e}nic gravitational contraction moves material
along the magnetic field lines, generating filamentary structures preferentially perpendicular to the
magnetic field \citep{nakamura08}.

\section{Formation models of filaments}\label{modreview}

Theoretical descriptions of filaments have focussed on several different kinds of possible filament
states: 1) equilibria, 2) collapsing and fragmenting systems that follow from unstable equilibria, 
3) equilibria undergoing considerable radial accretion and 4) highly dynamical systems for which
equilibrium descriptions do not apply. We refer the interested reader to a full discussion elsewhere
\citep[e.g. see Section 5 of][]{andre14} and go on to discuss formation mechanisms relevant to our work.

Early simulations have shown that
gas is rapidly compressed into a hierarchy of sheets and filaments, without the aid of gravity
\citep*{bastien83,porter94,vazquez94,padoan01}. Turbulent box simulations and colliding flows 
\citep[e.g.][]{maclow04,hennebelle08,federrath10,gomez14,moeckel15,smith14,kirk15}
produce filaments.
\cite{hennebelle13} demonstrated the formation of filaments through the velocity shear that
is common in magnetised turbulent media. Other authors have explained filaments as the stagnation
regions in turbulent media \citep{padoan01}. As discussed in the previous section, the formation of
filaments preferentially perpendicular to the magnetic field lines is possible in strongly magnetised 
clouds \citep{li10}.
\cite{andre14} note that the same 0.1\,pc filament width is measured for low-density, subcritical
filaments suggesting that this characteristic scale is set by the physical processes producing the 
filamentary structure. Furthermore, they note that at least in the case of diffuse gravitationally
unbound clouds (e.g. Polaris), gravity is unlikely to be involved. Large-scale compression
flows, turbulent or otherwise, provide a potential mechanism, but it is not clear why any of these
would produce filaments with a constant radius.

Filaments also form due to gravitational instabilites in self-gravitating sheets if the exciting modes 
are of sufficiently long wavelength \citep[e.g.][]{nakajima96,umekawa99}. Layers 
threaded by magnetic fields still fragment, but the growth of perturbations perpendicular to the 
magnetic field are suppressed when the thickness of the sheet exceeds the thermal pressure scale 
height according to the linear analysis of \citet*[][]{nagai98}. The perturbations parallel to the 
field are unaffected. Therefore filaments perpendicular to the magnetic field form first within the 
sheet, before they form any cores \citep[][]{inutsuka92}. \citet*[][]{vanloo14} 
showed numerically that filaments with properties similar to the observations indeed form by
gravitational instabilities, but that cores form simultaneously.

Recently, \cite[hereafter SGK14]{smith14} and \cite[hereafter KKPP15]{kirk15} investigated the 
formation and evolution of filaments in more detail. SGK14 examined the influence of different 
types of turbulence, keeping the initial mean density constant in simulations without magnetic fields.
Specifically they examined three turbulent initial conditions: solenoidal, compressive, and a 
natural mix of both - two-thirds solenoidal, one-third compressive. All were initialised with the
magnitude of the root-mean-square turbulent velocity normalised such that the kinetic and gravitational 
potential energies are equal at the start of the simulation. They used the moving mesh code AREPO and identified
and categorised simulated filaments from column density plots in the same manner as undertaken
for recent {\it Herschel} observations. They found that when fitted with a Plummer-like profile, the
simulated filaments are in excellent agreement with observations, with p $\approx$ 2.2, without
the need for magnetic support. They found an average FWHM of $\approx$ 0.3\,pc, 
when considering regions up to 1\,pc from the filament centre, in agreement with predictions for
accreting filaments. Constructing the fit using only the inner regions, as in {\it Herschel} observations,
they found a resulting FWHM of $\approx$ 0.2\,pc.

KKPP15 used the FLASH hydrodynamics code to perform numerical simulations of turbulent cluster-forming 
regions, varying density and magnetic field. They used HD and MHD simulations, initialised with a supersonic
($M \approx 6$) and super-Alfv{\'e}nic ($M_A \approx 2$) turbulent velocity field, chosen to match 
observations, and identified filaments in the resulting column density maps. They found magnetic
fields have a strong influence on the filamentary structure, tending to produce wider, less centrally peaked and 
more slowly evolving filaments than in the hydrodynamic case. They also found the magnetic field  is able to
suppress the fragmentation of cores, perhaps somewhat surprisingly with super-Alfv{\'e}nic motion
involved in the initial condition. Overall, they noted the filaments formed in their simulations have properties 
consistent with the observations they set out to reproduce, in terms of radial column density profile,
central density and inner flat radius.

Motivated by observed filamentary structure and the need to 
physically establish such structure within a hydrodynamic context for massive star feedback 
simulations, this study presents 2D fixed-boundary simulations and 3D free boundary simulations 
of an approach to the formation of the filaments: specifically, the action of the thermal 
instability (hereafter the TI) in a stationary quiescent diffuse molecular cloud initially 
in thermally unstable pressure equilibrium,  confined by its low density surroundings,
with only 10\% density variations seeded across the 100\,pc diameter cloud. 

\citet{parker53} was one of the first to suggest that condensation phenomena in molecular 
clouds could be a consequence of the instability that is a result of the balance between heating 
and cooling processes in a diffuse medium. \citet{field65} showed that the TI
can lead to the rapid growth of density pertubations from infinitesimal density variations, $\delta\rho$, to non-linear
amplitudes on a cooling time-scale, which for typical ISM conditions is short compared to the
dynamical time-scale. 
The TI develops an isobaric condensation mode and an acoustic mode, which - under ISM-conditions -
is mostly damped. The condensation mode's growth rate is independent of the wave length. 
However, since it is an isobaric mode, smaller pertubations will grow first \citep{burkert00}.
The signature of the TI is fragmentation and clumping as long as the sound crossing time is
smaller than the cooling time-scale. \citet{kritsuk02a,kritsuk02b} found that the TI can drive
turbulence in an otherwise quiescent medium, even continuously, if an episodic heating source
is available.

A number of authors have investigated analytically the effects of different mechanisms on the TI 
\citep*{birk00,nejad03,stiele06,fukue07,shadmehri09}.
Other groups have numerically investigated flow-driven molecular cloud formation
including the effects of the TI 
\citep*[e.g.][]{lim05,vazquez07,hennebelle08,heitsch09,ostriker10,vanloo10,inoue12}.
This numerical work has included magnetic fields, self-gravity and the TI and has identified the 
thermal and dynamical instabilities that are responsible for the rapid fragmentation
of the nascent cloud, largely through flow-driven scenarios.
Here we concentrate on the TI itself without any initial flow. As overdense regions appear in the 
molecular cloud, we continue each simulation with and without self-gravity in order to quantify the effect of
gravity on this large-scale initial stage of filament formation. We analyse the properties of these filamentary 
structures and compare them to the observational properties detailed above.

\section{Numerical Methods and initial conditions}\label{numerical}

\subsection{Numerical methods}

We present 2D and 3D, magneto-hydrodynamical (MHD) simulations of filament
formation from the diffuse atomic 
medium with and without self-gravity using the established astrophysical
code MG \citep{falle91}. The code employs an upwind, conservative shock-capturing scheme and is
able to employ multiple processors through parallelisation with the message passing interface
(MPI) library. MG uses piece-wise linear cell interpolation to solve the Eulerian equations of hydrodynamics. 
The Riemann problem is solved at cell interfaces to obtain the conserved fluxes for the time 
update. Integration in time proceeds according to a second-order accurate Godunov method
\citep{godunov59}. A Kurganov Tadmor \citep{kurg00} Riemann solver is used in this work.
Self-gravity is computed using a full-approximation multigrid to solve the Poisson equation.

The adaptive mesh refinement (AMR) method \citep{falle05} employs an unstructured grid 
approach. By default, the two coarsest levels (G0 and G1) 
cover the whole computational domain; finer grids need not do so. Refinement or 
derefinement is based on error. Where there are steep gradients of variable
magnitudes such as at filaments, flow boundaries or discontinuities, this automated
meshing strategy allows the mesh to be more refined than in more uniform areas.
Each level is generated from its predecessor by doubling the number of
computational grid cells in each spatial direction. This technique enables the generation
of fine grids in regions of high spatial and temporal variation, and conversely, relatively
coarse grids where the flow field is numerically smooth. Defragmentation of the AMR
grid in hardware memory is performed at every time-step, gaining further speed improvements
for negligible cost through reallocation of cells into consecutive memory locations.
The simulations presented below employed 7 or 8 levels of AMR. The coarsest level, G0, was set with a very small number of cells in order to
make the calculation of self-gravity as efficient as possible, specifically $4\times4\ (\times4)$ in 2D (3D). Thus,
the highest levels of grid resolution were either G6 with $256\times256\ (\times256)$ cells or
G7 with $512\times512\ (\times512)$ cells. Physical domain sizes and hence physical
resolutions varied as detailed below.

\subsection{Heating and cooling processes}\label{heatcool}

As the exploration of the evolution of thermal instability in the diffuse atomic medium under the influence
of magnetic fields and gravity is the aim of this paper, care has been taken to implement realistic equilibrium
heating and cooling, as it is the balance of these processes that is used to initialise the medium in 
the warm unstable phase and the combined effect of these that defines the evolution of the medium.
In the ISM, heating as defined by the coefficient $\Gamma$, varies with increasing density as
the starlight, soft X-ray and cosmic ray flux are attenuated by the high column density associated
with dense clouds. Because the exact form of the attenuation depends on details which remain
uncertain (e.g. the size and abundance of PAHs), the heating rate at T $\leq$ 10$^4$\,K is
similarly uncertain. In this work, as a first step, we have therefore assumed that $\Gamma = 2\times10^{-26}$ erg\,s$^{-1}$
(independent of density or temperature). For the low-temperature cooling ($\leq10^4$ K), we 
have followed the detailed prescription of \cite{koyama00}, fitted by \cite{koyama02}, corrected 
according to \cite{vazquez07}, namely
\begin{equation}
\begin{array}{ll}
\displaystyle{\frac{{\Lambda (T)}}{\Gamma } =} & \displaystyle{{10^7}\exp \left( {\frac{{ - 1.184 \times {{10}^5}}}{{T + 1000}}} \right)}\\
&\\
 & \displaystyle{+ 1.4 \times {10^{ - 2}}\sqrt T \exp \left( {\frac{{ - 92}}{T}} \right)}.
\end{array}
\end{equation}
The resulting thermal equilibrium pressure $P_{eq}$ and thermal equilibrium temperature $T_{eq}$, 
defined by the condition $\rho^2\Lambda=\rho\Gamma$, are shown in Fig \ref{fig1}(a) as a function
of density. Given the above forms of heating and cooling, it is possible to scale non-gravitational
simulations under the following transformation 
$\rho  \to \alpha \rho,~\Gamma  \to \alpha \Gamma,~t \to t / \alpha,~l \to l / \alpha $
where $\alpha$ is constant. This allows one to model different regions of the Galaxy which have different
heating rates \citep*[see e.g.][]{wolfire95,wolfire03}.

At temperatures above 10$^{4}$\,K we have followed the prescription of \cite{gnat12} who used CLOUDY 
10.00, enabling us to define cooling rates over the temperature range from 10\,K to 10$^8$\,K.
This has been implemented into MG as a lookup table for efficient computation. 
We do not expect such high temperatures in these simulations, but in order to enable stars and
their associated wind and SNe feedback to be introduced into these simulations in future, a consistent
approach from the outset has been used.

\subsection{Initial conditions}

The physical properties of our initial condition are motivated by the simplification of initial 
conditions and the need to avoid physical (or numerical) conditions which may pre-set a length 
scale in the simulation. For example, KKPP15 found that the injection of velocity on a particular 
forcing scale, as often used to initialise turbulent ISM conditions, can strongly affect the mean 
separation of filaments formed in the hydrodynamic case and to a lesser extent the magnetised 
case. We also take care to avoid numerical issues, e.g. instability-smoothing caused by AMR
derefinement, as discussed further below. In this way, we can examine the effect of the thermal
instability on diffuse medium evolution in isolation.

\begin{figure}
\centering
\includegraphics[width=80mm]{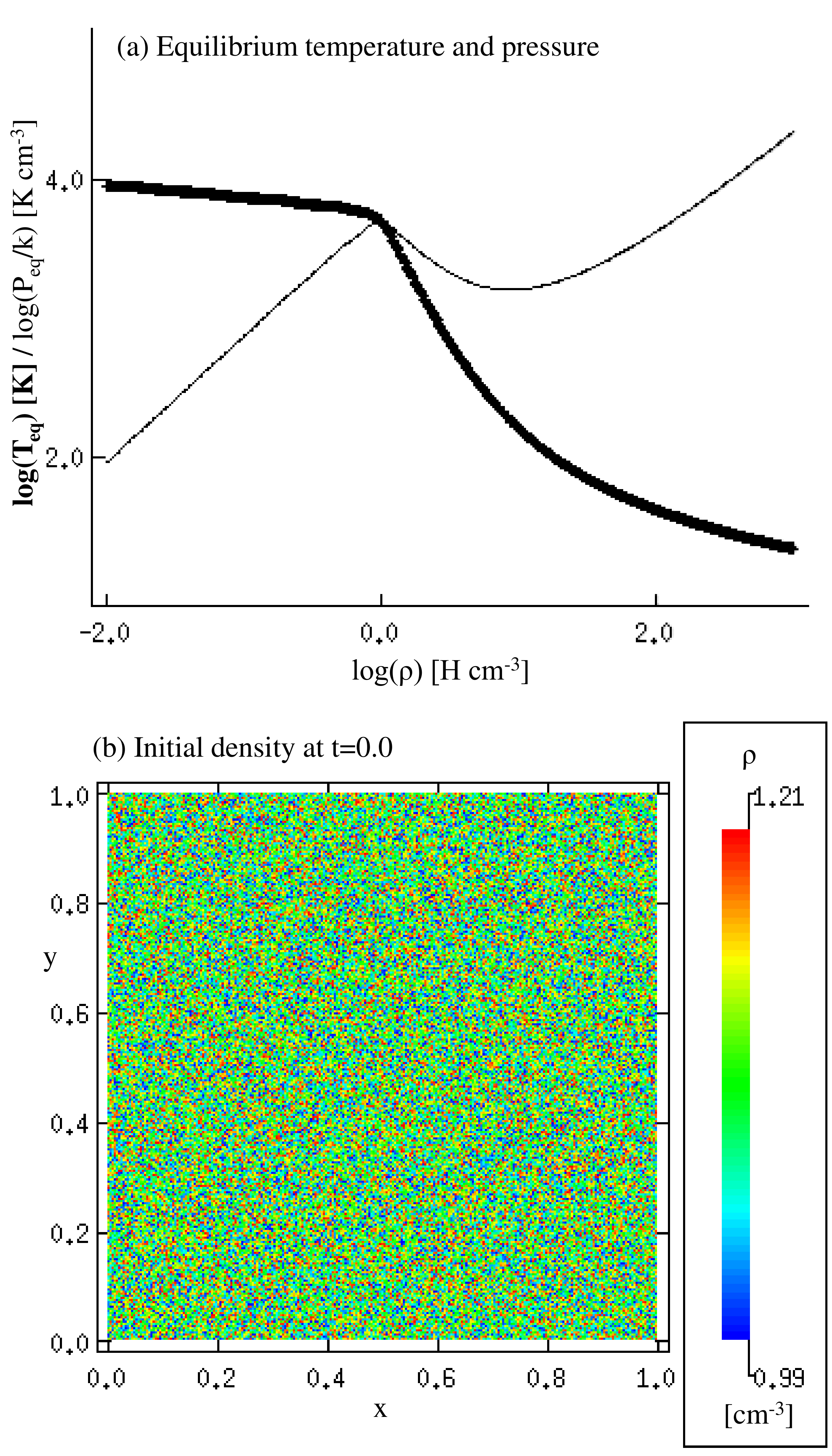}
\caption{(a) Thermal equilibrium pressure ($P_{eq}/k$ - thin line) and temperature 
($T_{eq}$ - thick line) vs. density for the cooling and heating functions given in Section
\ref{heatcool} and used to set the pressure and temperature in the initial condition.
(b) The initial density distribution across a 2D domain, showing 10\% density variations.
Length is scaled in units of 50\,pc.} 
\label{fig1}
\end{figure}

In our initial condition we set a number density of atomic hydrogen throughout the medium of n$_H$ = 1.1 cm$^{-3}$. 
Following \cite{field65}, we seed the domain only with random density variations - 10\% about this 
uniform initial density. We show the density distribution in Figure \ref{fig1}(b). Care is taken to avoid
repeating random number generation in multi-processor execution, by spooling through the random
number sequence independently according to processor number on each processor.
This initial density is at the lower end of the range of thermally unstable densities for the 
balance of heating and cooling functions used in this work, approximately n$_H$ = 1--7 cm$^{-3}$
defined by the region of negative gradient in the function of thermal pressure according to density
shown in Figure \ref{fig1}(a). Tests with a number of initial densities and $\pm$0.1\,cm$^{-3}$ 
variations across this range showed that a lower density triggers 
the formation of higher density structures on shorter timescales. Thus, the low value of initial density that 
we selected ensures that structure forms on the shortest timescales. From an evolutionary point 
of view, it would seem likely that mechanisms that form clouds from the ISM are likely to increase the 
density from typical ISM densities of 1 H\,cm$^{-3}$ or less into the lower end of the unstable phase 
first, although the passage of a shock may also jump the density straight to the upper end of the 
unstable regime, or higher, and several authors have investigated this as previously noted.
For the purposes of this investigation, we choose the former evolutionary process and consider the 
initial density that most quickly forms higher density structures whilst remaining on the equilibrium curve.
Initial pressure is set according to the unstable equilibrium of heating and cooling at 
P$_{eq}$/k = $4700\pm300$ K\,cm$^{-3}$ and results in an 
initial temperature T$_{eq}$ = $4300\pm700$\,K. The dependence of the equilibrium pressure on the
density of the medium is shown by the thin line in Figure \ref{fig1}(a). Equilibrium temperatures are
indicated by the thick line in Figure \ref{fig1}(a). The material is stationary.

In all cases, we follow the evolution for approximately a free-fall time of 
a diffuse medium at this density, where free-fall 
time $t_{ff}$ is defined as
\begin{equation}
{t_{ff}} = \sqrt {\frac{{3\pi }}{{32G\rho }}}
\end{equation}
and for these initial densities, $t_{ff} = 49.1 $\,Myrs.

For this work, we thread the domain with uniform $B$-field along the $x$ direction, i.e.
\textbf{B} = $B_0$ $\hat{\textbf{I}}_x$. We investigate three magnitudes of the magnetic
field, defined by a plasma $\beta$ = $\infty$, 1.0 and 0.1. In the case of $\beta = \infty$,
there is no magnetic field ($B_0$ = 0). In the moderate field case of $\beta = 1.0$, there is pressure equality,
i.e. the thermal pressure at n$_H$ = 1.1 cm$^{-3}$ is equal to the magnetic pressure at $t=0$, and $B_0$ = 1.15\,$\mu$G. 
In the strong field case of $\beta = 0.1$, the magnetic pressure is $10\times$ greater than 
the thermal pressure at $t=0$. The magnetic field strength is increased by a factor of $\sqrt{10}$ to 
$B_0$ = 3.63\,$\mu$G. Both cases
represent field strengths similar to mean Galactic values expected at an inner ($\sim$ 4\,kpc)
location and would be representative of the magnetic field conditions in which diffuse
clouds begin to condense and form higher density structures. Evolved molecular clouds have
been noted to have magnetic field strengths over 10$\times$ greater than these,
but it is not clear how these field strengths have been generated. In this work, we examine
whether thermal instability leading to filament formation under the influence of gravity alone
can intensify the magnetic field, or whether other influences are required, for example stellar
feedback generating super-sonic, super-Alfv{\'e}nic motions. 

We considered three separate Scenarios, employing 2D $XY$ and 3D $XYZ$ Cartesian grids, in 
order to examine the evolution of the thermal instability:

\subsubsection{Scenario 1 - a segment of a larger cloud (2D)}

In this Scenario, 2D simulations employed fixed boundary conditions that enforced the initial condition 
at all boundaries in order to represent a segment of a larger molecular cloud. Periodic boundary
conditions were avoided in order to allow the use of self-gravity. Three simulations
were performed without self-gravity for the three magnetic field cases $\beta = \infty$, 1.0 and 
0.1, and then repeated with self-gravity. In every simulation, the domain was filled with the initial 
condition detailed above, as shown in Figure \ref{fig1}(a). The physical domain size was 50\,pc$^{2}$ throughout, with G0 
containing $20\times20$ cells and 5 or 6 levels of AMR initially, resulting in a resolution 0.15625\,pc 
on G4 or 0.078125\,pc on G5 ($640\times640$). In all these simulations, the AMR capability was disabled and 5 or 6 
complete grid levels were simulated as tests showed that during the initial evolution of the medium, the 
density variations seeded in the initial condition initially smoothed out before condensations appeared
across the cloud. Hence the AMR grid would by default completely derefine, suppressing condensations. 
Further tests, not shown, were also carried out with both half-physical-size domains and double 
resolution domains, hence twice the physical resolution on the finest grid level. These tests converged 
with the lower physical resolution simulations, in terms of the numbers of structures, their size and 
separation.

\subsubsection{Scenario 2 - a slice of an infinite cylinder (2D)}

In this Scenario, a circular stationary diffuse cloud of radius 50\,pc was placed at the origin (0,0) in a domain
of physical extent $\pm75$\,pc in both directions ($150$\,pc square), surrounded by a
lower density stationary medium. The minimum 25\,pc buffer around the cloud in all directions 
avoided any boundary-related numerical effects. Free-flow boundary conditions were employed on all 
boundaries. Again, three simulations were performed without self-gravity for the three magnetic field 
cases, and then repeated with self-gravity. The same initial condition as previously noted was 
adopted inside the cloud and the cloud was assumed to have a definite edge, i.e. no smoothing 
between the cloud and its surroundings was adopted. The surrounding medium was set with a density 
of 0.1\,H\,cm$^{-3}$ in pressure equilibrium with the cloud and hence at a high temperature. If
allowed to evolve this surrounding medium would cool rapidly. As we are not interested in the 
evolution of the surrounding medium and its pressure is simply defined in order to confine the 
cloud but not affect its internal evolution, heating and cooling was disabled in the surrounding 
medium. The same effect was achieved in Scenario 1 by the use of fixed boundary conditions. 

\subsubsection{Scenario 3 - a spherical cloud (3D)}

In this Scenario, the same stationary diffuse cloud of radius 50\,pc was surrounded by a stationary medium 
in a $150$\,pc cube 3D domain. The same initial conditions as previously were 
adopted, both inside the cloud, resulting in a total cloud mass of $\sim17,000$\,M$_\odot$, and
also in the surrounding medium outside the cloud. In
this Scenario, G0 contained $4\times4\times4$ cells and the simulation employed 8 levels of AMR, 
equivalent to a $512\times512\times512$ grid and a finest resolution of 0.29\,pc on a side. A low
resolution G0 was adopted in order to make the calculation of self-gravity more efficient. Levels G0 
to G5 were initially complete in order to avoid the numerical AMR instability-smoothing issues 
discussed above. Levels G6 and G7 were allowed to refine or derefine according to the physical
conditions in the domain. Tests again showed that this approach and resolution captured the true physical 
evolution of the cloud and avoided coarse grid scales affecting the formation and evolution of 
higher density structures. No symmetry constraints were imposed on the simulation and all quadrants 
were calculated so that asymmetric structure can develop freely throughout the cloud. Free-flow
boundary conditions were used at all boundaries.

\subsection{Neglected processes and simplifications}

\begin{figure*}
\centering
\includegraphics[width=170mm]{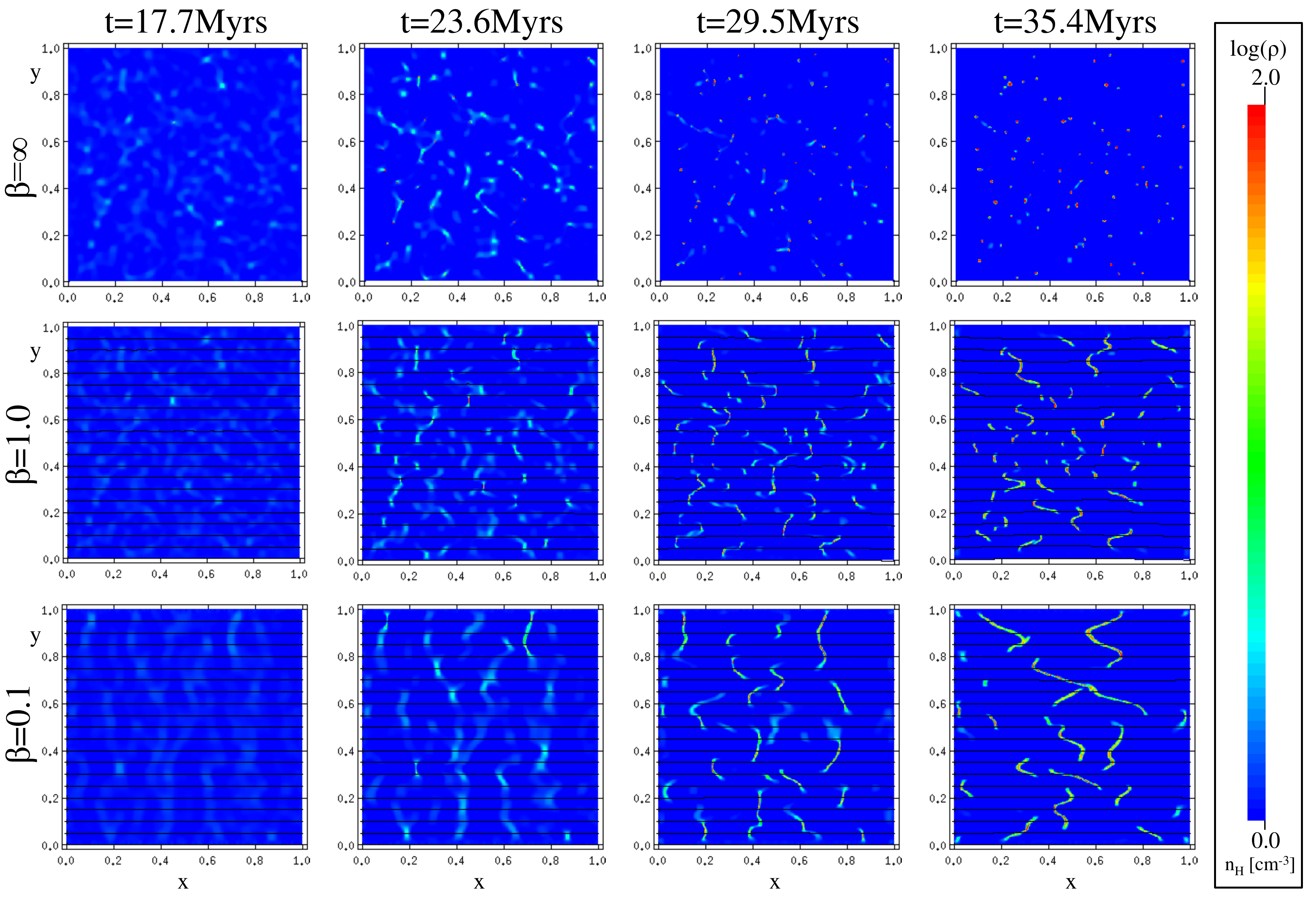}
\caption{Scenario 1 - 2D simulations of segments of a larger cloud, {\it without} self-gravity. Each plot
shows the logarithmic mass density. 
Across the rows, the three different magnetic field cases are presented. Magnetic field 
lines are indicated where appropriate. Length is scaled in units of 50\,pc.} 
\label{fig2}
\end{figure*}

This work is the first step in developing realistic molecular cloud conditions for stellar feedback
simulations, such as we have performed previously with predefined cloud conditions
\citep{rogers13,rogers14}. We have accounted for radiative heating and cooling, gravity and 
magnetic fields. We have necessarily still made a number of simplifications and approximations.

We model equilibrium cooling only and neglect the role of molecular cooling, including carbon 
monoxide (CO). Without a full treatment of heating according to column
density and shielding to allow the formation of CO, it is difficult to justify the inclusion of any
CO effects into the cooling curve. That said, we have performed a small number of tests with
just such an amendment to the low temperature cooling, as used by \cite{rogers13,rogers14}.
We have found that the increased cooling introduced by CO at lower temperature allows the clumps and filaments to cool 
further (with associated increased density) to temperatures on the order of 10-15\,K. 
With regard to this work, it should 
be noted that the densities we find are likely to be at the lower end of the range of observed clump and
filament densities. We leave a more complete treatment of the role of CO to a future work.
Internal photoionisation does not play a role as no stars have formed yet in these molecular clouds.
In a forthcoming work, we examine the introduction of stars and their wind and SNe feedback into
the cloud.



\section{Results}\label{results}

In this section we present our results. Firstly, we examine the evolution of the Scenario 1 2D box 
simulations (Section \ref{res-case1}). Next we focus upon the evolution of the 2D diffuse cloud 
simulations up to a free-fall time in Section \ref{res-case2}. Finally, in Section \ref{res-case3} we present the evolution
of the 3D diffuse cloud simulations, presenting slices through the simulation domain parallel and 
perpendicular to the imposed field, as well as collapsing the simulation domain perpendicular 
to the imposed field. In the following Section \ref{analysis} we analyse and discuss the 
properties of the individual clumps and filaments.

In the following sub-sections, we refer to the formation of clumps from the
diffuse cloud. We separate the terminology in this fashion to achieve clarity for the reader, although
it should be noted that our definitions do not track identically with those typical of other authors, e.g.
Table 1 of \cite{bergin07}. Our "clumps" are similar to Bergin \& Tafalla's clouds, although our
velocity dispersions are lower, for reasons investigated in the next Section.

\subsection{Scenario 1 - a segment of a larger diffuse cloud}\label{res-case1}

\begin{figure*}
\centering
\includegraphics[width=170mm]{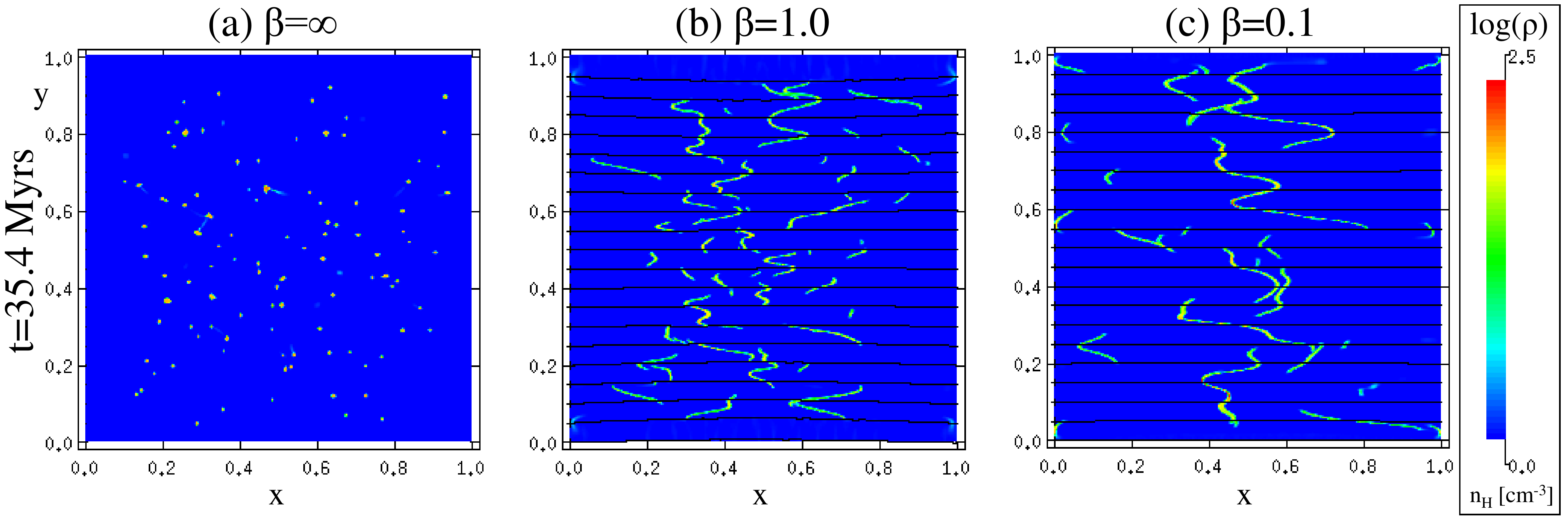}
\caption{Scenario 1 - 2D simulations of segments of a larger diffuse cloud, {\it with} self-gravity. Each plot
shows the logarithmic mass density. Across the columns, the three different magnetic field cases
are presented, all at the same snapshot time in the simulation of 35.4\,Myrs.
Magnetic field lines are indicated where appropriate. Length is scaled in units of 50\,pc.} 
\label{fig3}
\end{figure*}

We begin by showing the time evolution of the three {\it diffuse cloud segment} simulations without 
self-gravity in Scenario 1 (Fig. \ref{fig2}). In the first 10\,Myrs, all cases 
evolve to smooth the initial density pertubations shown in Fig. \ref{fig1}(a). Material in 
isolated unstable thermal equilibrium can then evolve into one of two stable states -- either it contracts
and cools into the cold state, or warms and expands into the warm state.
We observe regions of the domain undergoing precisely this -- some begin to cool and contract to higher density, 
others begin to warm and reduce in density. The flow of material from warm to cool regions then
accentuates the density inhomogeneities further. In the absence of magnetic field and gravity, the distribution 
and growth rates of the high density regions is controlled by the TI ($\beta=\infty$, top row Fig. \ref{fig2}). 
The initial smoothing has no preferred direction and condensations appear 
after 18\,Myrs with densities a few tens of times higher than the initial condition across the domain. 
By 24\,Myrs, it's clear that these ``clumps" are not growing at the same rate; some 
are already far denser than others. Mass flows 
onto the clumps from all directions equally and hence the clumps are randomly distributed across the domain. 
Lower density linear structures interconnect a number of the clumps. These structures fit the 
definition of filaments, but are in fact transitory structures of material flowing towards  
higher density clumps. An examination of the velocities in the domain confirms
this - the clumps are stationary, with material moving onto them equally from all directions, 
including motion toward clumps along the filamentary structures. The spacing of the clumps is 
approximately 5-10\,pc. By 29.5\,Myrs (third column), the number of such linear 
structures has greatly reduced, although some do persist to 35\,Myrs and beyond.
The clumps have a range of densities from a few hundred H\,cm$^{-3}$ up to 
1000\,H\,cm$^{-3}$ and are distributed evenly across the domain. The TI has enabled the initial density 
inhomogeneities to contract into high density clumps on a time-scale considerably shorter than
the dynamical free-fall time of the cloud, according to the growth described by \cite{field65}.
The lowest temperatures correspond to the equilibrium conditions observed in the highest density
regions, around 30\,K. We discuss the properties of the clumps, including temperature and velocity 
profiles in more detail in Section \ref{analysis}.

The TI generates motions that are sub-sonic outside the cool dense structures and sub-Alfv{\'e}nic
everywhere in the domain. They are on the order of 5\,km\,s$^{-1}$ towards the filaments. 
The internal motions inside the filaments are also sub-sonic on the order of 0.5 - 0.6\,km\,s$^{-1}$.
In the transition regions, where material reaches the filament, trans-sonic motions are
briefly observed. With $\beta=1.0$ (Fig. \ref{fig2}, second row), the magnetic field preordains  
the direction for the flow of material - along the field lines only with this velocity range (indicated by 
lines across the domain). Condensations that have appeared by 18\,Myrs show only minor differences 
to those formed with $\beta=\infty$ at that time. By 23.6\,Myrs the linear structures seen previously 
are now stationary filaments that grow in density at the same rate as the clumps did previously, gaining
material from flow along the field lines. By 29.5\,Myrs, their nature as persistent and stationary filamentary 
structures is clear. The highest density filaments are predominantly perpendicular to the magnetic field 
direction. A number of lower density linear structures `feeding' the high density filaments are now 
apparent, which are more likely to be parallel to the field direction, in agreement with the direction of 
sub-sonic, sub-Alfv{\'e}nic motion defined by the magnetic field.
There are also a number of objects that are more clump-like. They are somewhat extended perpendicular
to the field but not enough to be defined as a filament (according to our definition,
see Section \ref{obsreview}). Both the filaments and the clumps show a similar range of densities,
up to a few hundred H\,cm$^{-3}$ -- lower than $\beta=\infty$. Another difference from $\beta=\infty$ is that
between 29.5\,Myrs and 35.4\,Myrs, the filaments now move considerably more, still along the field
lines, connecting and forming longer, more curled filaments that have similar densities and widths, 
but are now considerably longer. Many of the clumps seen at 29.5\,Myrs have now been
absorbed into these filamentary structures.

\begin{figure*}
\centering
\includegraphics[width=174mm]{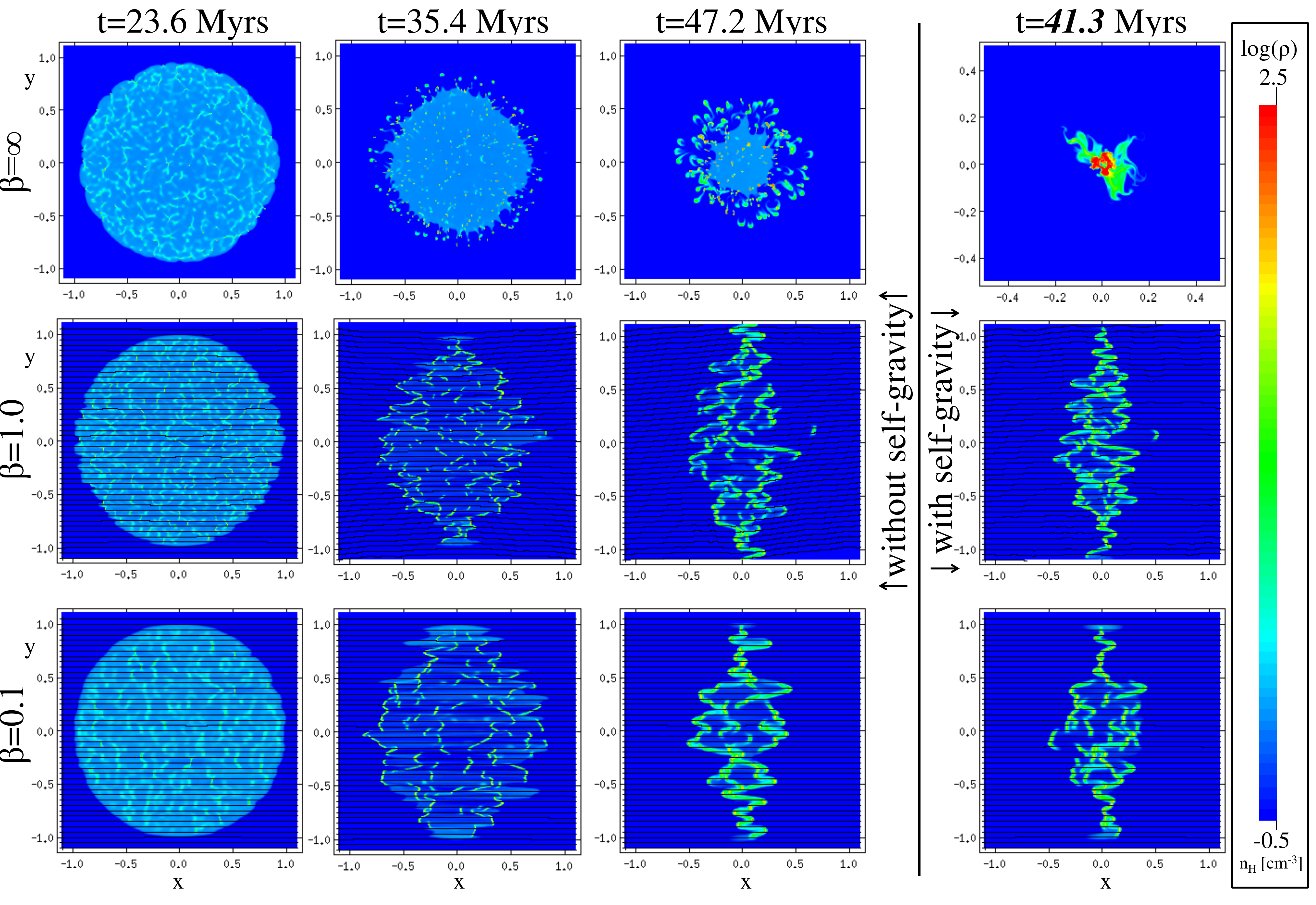}
\caption{Scenario 2 - 2D simulations of slices perpendicular to the major axis of a larger cylindrical 
cloud, in columns 1-3 without self-gravity and in column 4 with self-gravity. Each plot shows 
the logarithmic mass density. Across columns 1-3, time evolution of the simulations without gravity is 
presented. Across the rows, the three different magnetic field cases are presented. 
Magnetic field lines are indicated where appropriate. Length is scaled in units of 50\,pc.} 
\label{fig4}
\end{figure*}

With $\beta=0.1$ (Fig. \ref{fig2}, third row), the magnetic field now dominates the evolution
and the only structures that form are filamentary. Rather than growing in length at late times, these
filaments initially cool and condense out of the smoothed initial condition as long, 
high-density structures with high aspect ratios. 
For the first 30\,Myrs, they are almost exclusively perpendicular to the magnetic field. 
In the final plot (fourth column), after 35.4\,Myrs of evolution, the filaments are now skewed across 
the domain, a consequence of the initial random density distribution.
The magnetic field remains unaffected by this motion. 

In all three cases after 35.4\,Myrs, large-scale movement occurs across the
entire domain. The effect of self-gravity on this movement is important and so we
go on to consider the repeat of these simulations with self-gravity, as shown in Fig. \ref{fig3}.
With $\beta=\infty$, the major difference is in the range of clump densities formed during the 
evolution. Higher density clumps than previously are formed on the same time-scale.
In both $\beta=1.0$ and $\beta=0.1$ cases, the major difference as expected 
is at late-time (t=35.4\,Myrs $\approx$ 0.7\,t$_{ff}$). Gravitational attraction has resulted 
in the movement of the filaments
towards one another and the centre of the domain, albeit with velocities that are still 
sub-sonic and sub-Alfv{\'e}nic and hence the filaments appear to crowd around $x=0.5$.

We have included this Scenario in order to study and document the action of the 
TI in isolation, but it is questionable how representative these simulations are of a cloud 
segment after 30\,Myrs. In the next two sections, we model a cloud and surroundings in 
order to address this point more fully.

\subsection{Scenario 2 - slices of cylindrical clouds}\label{res-case2}

In Fig. \ref{fig4} we show the time evolution of the three magnetic field cases with 
and without the effect of self-gravity for Scenario 2, representing a slice through a cylindrical
cloud perpendicular to the major axis. With $\beta=\infty$, the cloud evolves without any magnetic field and the natural
action of the TI can be observed. By 23.6\,Myrs, cold condensations have formed across 
the cloud. As the domain is now considerably larger (shown in the figure is 100\,pc$\times$100\,pc), 
many more condensations are apparent in the domain. As before, they evolve to form small, 
cold, high density clumps by 35.4\,Myrs. Motion caused by the TI has caused the contraction 
due to the pressure loss within the cloud. Self-gravity does not play a role in this simulation.
with by late-time (47.2\,Myrs), a smaller number of more dense clumps are now 
apparent, as well as clumps on the outer edge of the cloud losing mass towards the centre of the cloud.
Further detailed investigation of the simulation, in particular the velocity in and around the clumps, reveals that they 
are falling radially inwards at a lower speed than their lower density surroundings. In the frame 
of reference of the clump, lower density material is therefore flowing past the clump and entraining 
clump material into the faster flow. The result is the formation of tails behind the clumps and directed
radially inwards, as seen in the figure.
Given that around the edges of the cloud, the TI will generate motion toward the centre of the cloud
accentuated by the 2D numerical approach, the authenticity of this collapse will be studied in
more detail in the next sub-section, considering a 3D simulation.

\begin{figure*}
\centering
\includegraphics[width=175mm]{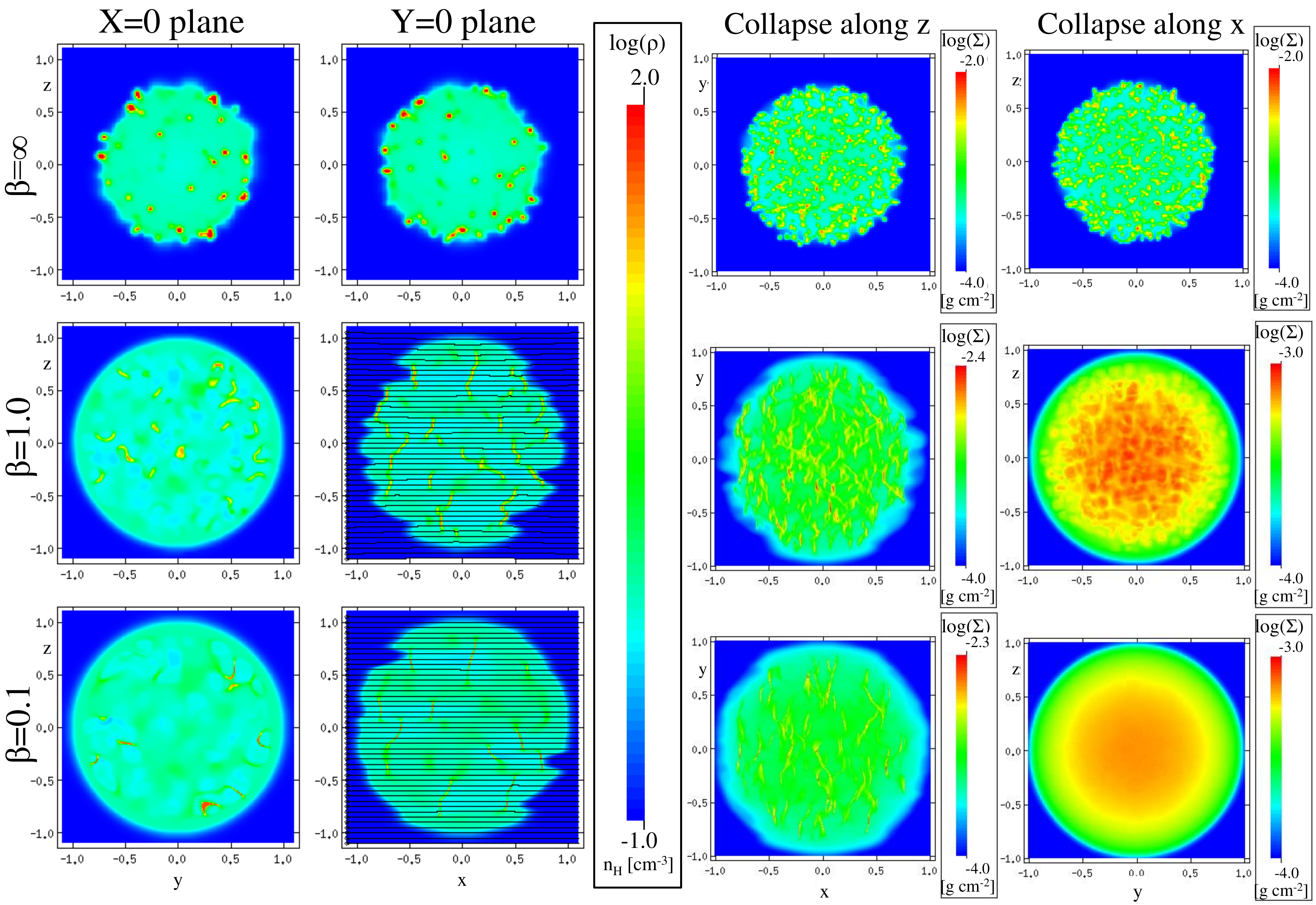}
\caption{3D simulations {\it without} self-gravity at t=35.4\,Myrs. Columns 1 and 2 show the logarithmic mass density
on planar slices through the domain at $x=0.0$ perpendicular to the field and $y=0.0$ parallel 
to the field respectively. Columns 3 and 4 show the logarithmic column density by projection along the $x$ 
axis onto a $y$-$z$ plane and along the $z$ axis onto a $x$-$y$ plane respectively. Across the rows, 
the three different magnetic field cases are presented. Magnetic field lines are indicated 
where appropriate. Length is scaled in units of 50\,pc.} 
\label{fig5}
\end{figure*}

With the introduction of self-gravity to the $\beta=\infty$ field case, the cloud  
contracts under the effect of both the TI and gravity to a high density core 
by 41.3\,Myrs, consistent with the free-fall time of 49.2\,Myrs. Approximately 
17,000\,M$_{\odot}$ is now contained within a radius of ~5\,pc. Structures formed by 
the TI during the preceding evolution have been destroyed by the gravitational collapse of the cloud.

With $\beta=1.0$, structures form across the cloud in the first 25\,Myrs in a similar manner to
those observed in the $\beta=\infty$ field case. The two clouds are almost indistinguishable. After
25\,Myrs, the evolution of these two clouds diverges. With magnetic pressure equal to
thermal pressure, filaments and clumps again form across the cloud. The radial contraction
of the cloud seen without magnetic field now manifests itself as contraction along the field
lines only. An elliptical cloud forms, supported in the direction perpendicular to the magnetic 
field by the magnetic pressure. The cloud evolves
towards becoming a bundle of filaments. By 47.2\,Myrs, many of the smaller filaments have
now inter-connected and formed longer, more curled structures. These have also absorbed
many of the clumps that had evolved up to this point. Large-scale
motion is directed towards the vertical axis of the cloud, with larger velocities at late-time. 
In this simulation with self-gravity (as seen in the 4th column of Fig \ref{fig4}), the filaments 
survive the gravitational collapse. The filaments eventually gather towards the centre of the
cloud, moving along the field lines, with the magnetic field providing support against both 
the TI and the gravitational collapse of the cloud.

In the simulation with $\beta=0.1$, the effect of a strong magnetic field can be seen even at comparatively early
time in the evolution of the cloud. The condensations forming out of the low density cloud are
already perpendicular to the field, indicating motion of material is very strongly confined along
field lines. By 35.4\,Myrs the cloud is dominated by filaments predominantly perpendicular
to the imposed field. The TI-driven collapse along the field lines leads more rapidly towards the ordered
extended filament situation seen in the $\beta=1.0$ case, with sub-filaments that contain 
distinct substructures in both density and velocity. With gravity, the collapse is even more rapid 
towards this conclusion. 

\begin{figure*}
\centering
\includegraphics[width=170mm]{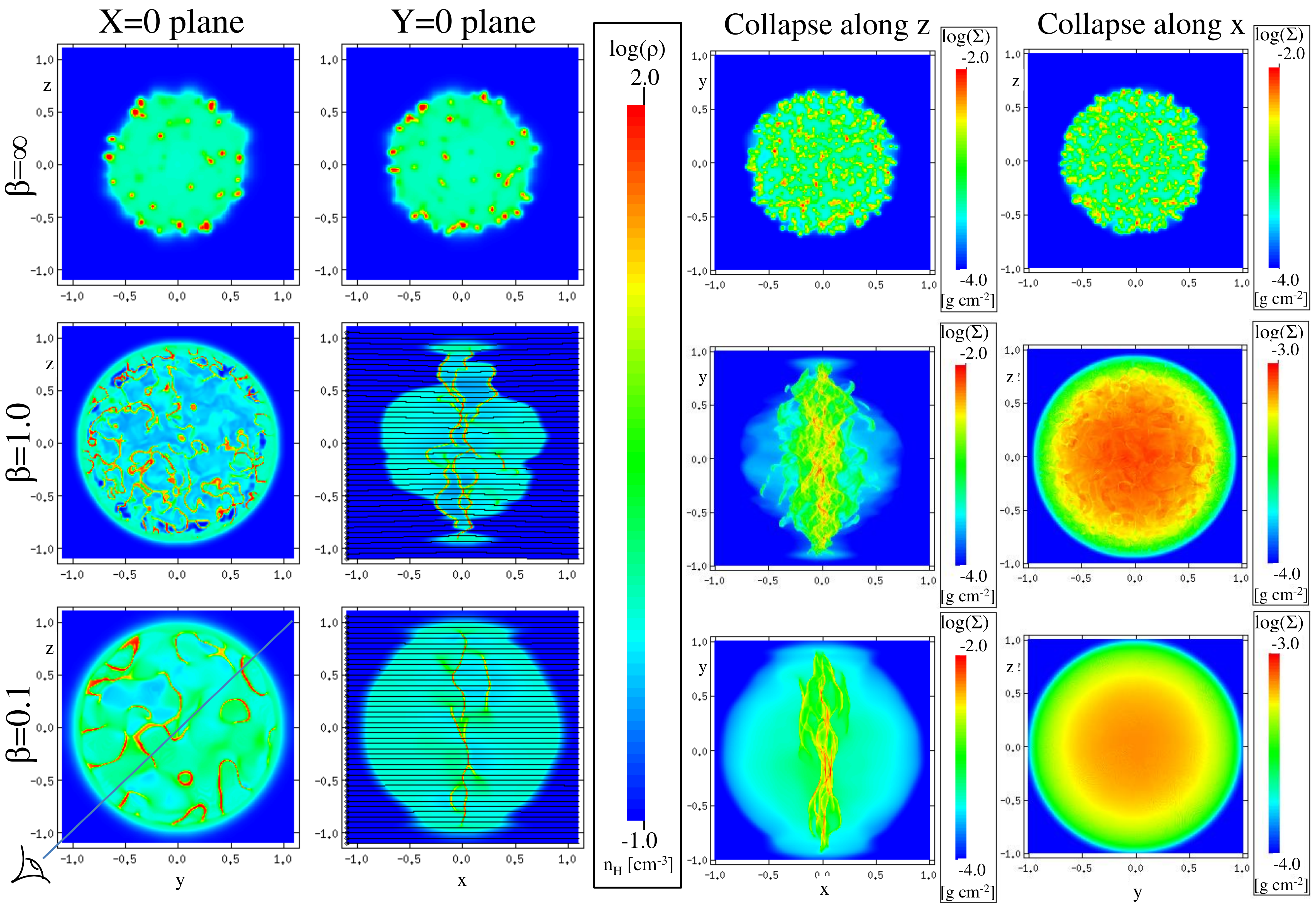}
\caption{3D simulations {\it with} self-gravity at t=35.4\,Myrs. Columns 1 and 2 show the logarithmic mass density 
on planar slices through the domain at $x=0.0$ perpendicular to the field and $y=0.0$ parallel 
to the field respectively. Columns 3 and 4 show the logarithmic column density by projection along the $x$ 
axis onto a $y$-$z$ plane and along the $z$ axis onto a $x$-$y$ plane respectively. Across the rows, 
the three different magnetic field cases are presented. Magnetic field lines are indicated 
where appropriate. Length is scaled in units of 50\,pc.} 
\label{fig6}
\end{figure*}

\subsection{Scenario 3 - a spherical cloud}\label{res-case3}

We now consider our 3D simulations of spherical clouds. In Fig \ref{fig5}, we show slices 
through the domain parallel and perpendicular to the magnetic field, as well as projected 
column densities in order to gain the most insight into the evolution of these molecular clouds.
Figures in the previous two Scenarios have shown how the clouds evolve to form filaments, 
and so in this section we choose to show the cloud at a particular instance in its evolution
-- 35.4\,Myrs -- once structure has formed but less than t$_{ff}$. The evolution up to this
point has been illustrated in the previous two Scenarios. Where there are deviations from this
evolution, we note those in our description.

With $\beta=\infty$ (Fig. \ref{fig5}, top row), the TI triggers the now familiar formation of cold, dense
clumps across the molecular cloud, predominantly towards the edge of the cloud, the properties of
which we will discuss in detail in the next section. Planes through the simulation domain
at $x=0$ and $y=0$ are qualitatively the same in terms of the major characteristics, e.g. 
overall cloud dimensions, clump distribution, clump density and interclump conditions. 
The quantitative differences are a consequence of the random initial conditions.
Collapse of the datacube along either the $z$ or $x$ axis also results in indistinguishable column
density projections. It is worth noting that the spread of clumps across the cloud in projected
column density is roughly even, i.e. there is no increased population density towards the centre of the cloud. This
indicates the clump formation, at least in this Scenario without self-gravity, is predominantly around
the edges of the cloud rather than uniformly across the cloud, which would result in increased
clump population density towards the centre of the cloud. Simulated in 3D, it would appear that collapse of the
cloud under the influence of the TI alone is less pronounced. Given that in 2D this is caused
by velocities developing radially inwards toward the centre of the cloud, in 3D a wider range
of directions of velocity can develop and the rapid collapse observed in 2D is revealed as an 
artifact of 2D modelling. Whilst there are no indications of clump ablation in this simulation at
this time, we find similar but less pronounced effects around the edge of the cloud by t$_{ff}$,
$\sim$53\,Myrs.

With a magnetic field with $\beta=1.0$ (Fig. \ref{fig5}, second row), multiple filaments appear to form
uniformly across the cloud, perpendicular to the field direction on both a plane at $y=0$ and 
in projection collapsing the density distribution along the z axis. These filaments persist rather than merge.  
The question of whether these filaments are actually individual filaments or corrugated
interconnecting sheets seen in projection can now be addressed. Inspection of the 3D 
simulation would indicate that for this field strength, without self-gravity, filaments form 
separately and eventually merge as more material moves out of the thermally unstable
state, creating first filaments, then interconnected `corrugated' sheets, with density 
varying across the sheet creating structure in the sheet and in projection. A line of sight
across the sheet would see several filaments, that are in fact interconnected (see the
illustrative line of sight in Fig. \ref{fig6} for the $\beta=0.1$ case on the $x=0$ plane.)
The preceding filaments persist for a relatively long period of the evolution - from 25\,Myrs 
to 35\,Myrs as noted previously.

\begin{figure*}
\centering
\includegraphics[width=175mm]{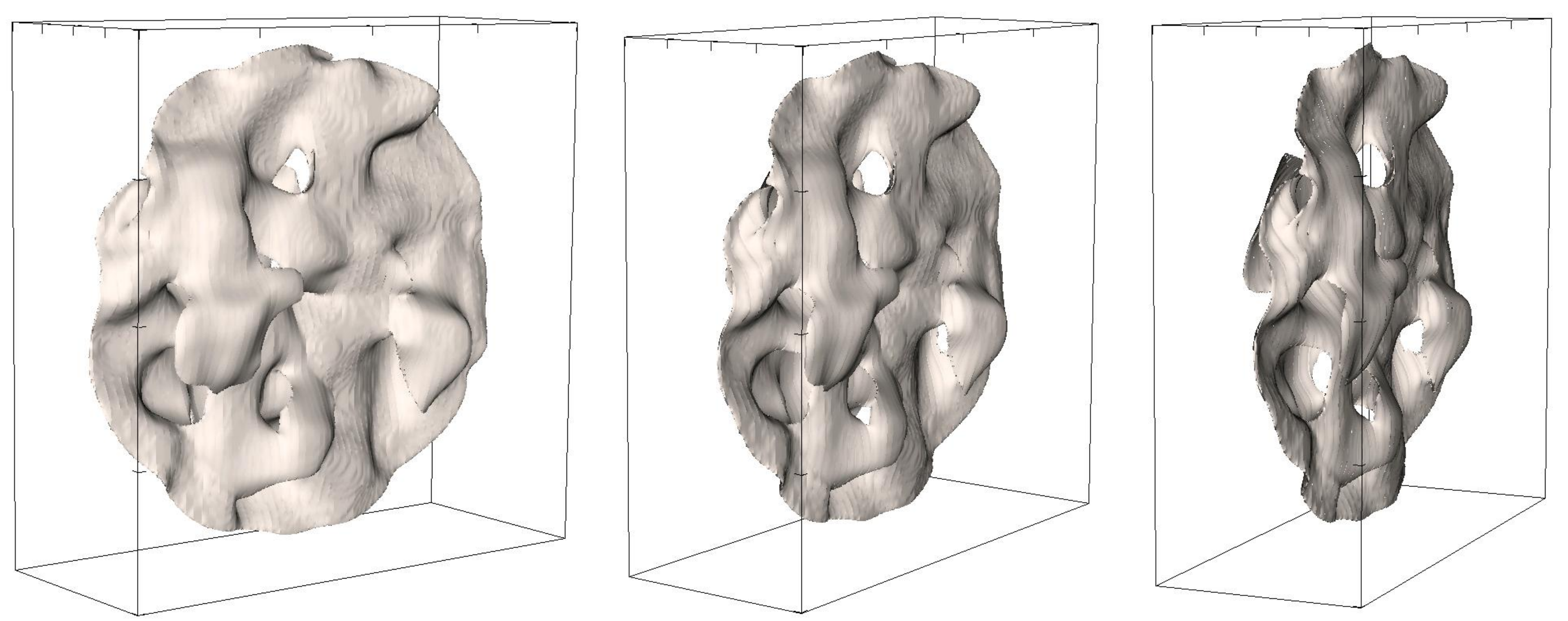}
\caption{Three different views of the same 3D isosurface of constant density (10 cm$^{-3}$) 
illustrating the corrugated nature of the single sheet formed in the $\beta=0.1$ case with 
self-gravity at 35.4\,Myrs. Visualisation created using the VisIt software \citep{visit}. 
Length is scaled in units of 50\,pc.} 
\label{fig7}
\end{figure*}

Projecting perpendicular to the magnetic field to create a collapsed plane parallel to the
field (third column), the cloud appears entirely filamentary. Column densities in the
filaments are on the order 10$^{-3}$ -- 10$^{-2}$\,g\,cm$^{-2}$. This is in excellent
agreement with range of column densities derived from {\it Herschel} data in a portion
of the Polaris flare translucent cloud \cite{andre10} and also central column densities
in the B213/B211 filament in Taurus \cite{palm13}. The nature of the velocity 
dispersions in these filaments generated by the TI is on the order of 0.5 - 1.0\,km\,s$^{-1}$.
The velocity variation across the corrugated sheets shows dispersions around
separate velocity components for different filaments. Separated velocity components
have previously been interpreted as evidence for separated filamentary structures, 
rather than corrugated sheets. However, it is now clear that corrugated sheets, with velocity variations
across the sheets of several km\,s$^{-1}$ (up to $10\times$ larger than the non-thermal velocity
dispersions) and density variations caused by the TI leading to a filamentary appearance in 
projection, can also generate separated velocity components, with dispersion around these
components. This indicates that separated velocity components should not
be used to differentiate between filamentary structures and corrugated sheets seen in projection.

The major axis of the filamentary structures are perpendicular to the 
magnetic field, although corrugations in the sheets show structures which are parallel 
to the field in places. Unlike in the previous Scenario (c.f. Fig \ref{fig3}), there are no indications 
of isolated clumps in this cloud. As in the previous Scenario, the highest density filaments are exclusively
perpendicular to the magnetic field. The lower density filamentary structures interconnecting
these high-density filaments are again regions still undergoing TI-driven condensation
flowing onto the higher-density filaments. 
Projecting along the magnetic field to create a collapsed plane perpendicular to the
field (fourth column), shows structure apparent in the cloud, but it is not clear that it is at all filamentary. A cloud
seen along a line of sight perpendicular to the field (third column) can appear very
different from the same cloud seen along a line of sight parallel to the field (fourth column).

With $\beta=0.1$ (Fig. \ref{fig5}, third row), fewer filaments form and they are exclusively perpendicular to the 
magnetic field. The average filament spacing would appear to be larger than in the 
$\beta=1.0$ case. Seen in projected column density, the cloud appears 
filamentary when the line of sight is perpendicular to the field and {\it uniform} when 
the line of sight is parallel to the field. Comparable column densities to the $\beta=1.0$
case and observations are produced perpendicular to the field. Parallel to the field, 
column densities are $10\times$ lower. Clearly $\beta$ is low enough to enforce no
movement of material across field lines -- due to entirely sub-Alfv{\'e}nic velocities. 
We are not aware that such strongly contrasting
numerical outcomes have been seen before when considering projected column densities.
KKPP15 shows filaments appear filamentary when projected in all three directions, although
they only show column density distribution projected along the $x$ axis in their simulations.

In Fig. \ref{fig6} we show the evolution of the spherical cloud including the effect of self-gravity.
Without a magnetic field, the cloud undergoes gravitational collapse and shrinks as shown in 
the top row of Fig. \ref{fig6}. Unexpectedly, the projected column density both along $x$ and $z$
reveals a few linear filamentary structures connecting the high density clumps. Although there are
some linear structures visible in the $y=0$ plane, it is more than likely that these are in fact a 
projection effect. With $\beta=1.0$ (Fig. \ref{fig6}, second row), filaments that initially formed all across the cloud in the 
simulation without gravity, have now formed predominantly around $x$=0 and are merging to 
form a complex bundle of filaments seen in projection across the field (third column). In reality,
this is the projection of a number of interconnecting sheets located roughly parallel to a plane 
at $x$=0, rather than individual filaments. Across the field lines, the magnetic field has supported 
the cloud from TI-driven and gravitational collapse. In the $\beta=0.1$ case (Fig. \ref{fig6}, third row), 
by 35.4\,Myrs the cloud has formed a single sheet down the centre of the cloud.
The slice plane at $x=0$ (Column 1) highlights the corrugated nature
of the sheet. This sheet is notably less corrugated than in the $\beta=1.0$ case - there are fewer
intersections of the sheets shown by the red high density on the $x$=0 plane.
Parallel to the field, the slice plane at $y=0$ highlights the single filamentary
nature of this cloud. 3D isosurfaces of constant density best illustrate the corrugated 
nature of this sheet, as shown in Fig. \ref{fig7}.
Both $\beta$=1.0 and 0.1 simulations with self-gravity are more
filamentary than in the simulations without self-gravity (c.f. Fig \ref{fig5}). 
Collapsing along the z-axis to examine the column density, 
a tightly bound bundle of filaments is the apparent manifestation of the corrugated sheet 
formed by the thermal instability. Peak column densities are comparable to previously.
In projection along the field direction, the same uniform cloud 
is observed as in the simulation without self-gravity, with no internal structure, but column
densities $10\times$ lower. The dominance
of the magnetic field has guaranteed that material has only moved along the field lines, resulting
in the smooth projection of an apparently uniform density sphere onto this plane.
We go on now to examine whether the apparently filamentary structures formed in Scenario 3
with self-gravity in particular show any resemblance to observed structure.

\section{Clump and filament analysis}\label{analysis}






\subsection{Clumps}

Clumps are formed throughout the cloud in all three Scenarios with the $\beta=\infty$ field 
case and also along with filaments in the $\beta=1.0$ field case. In this
section, we present their properties.
We remind the reader that we use ``clump" by choice in order to achieve 
clarity from the diffuse ``cloud" initial condition and it should be noted that this definition does 
not track identically with those typical of other authors, e.g. Table 1 of \cite{bergin07}.

\begin{figure}
\centering
\includegraphics[width=78mm]{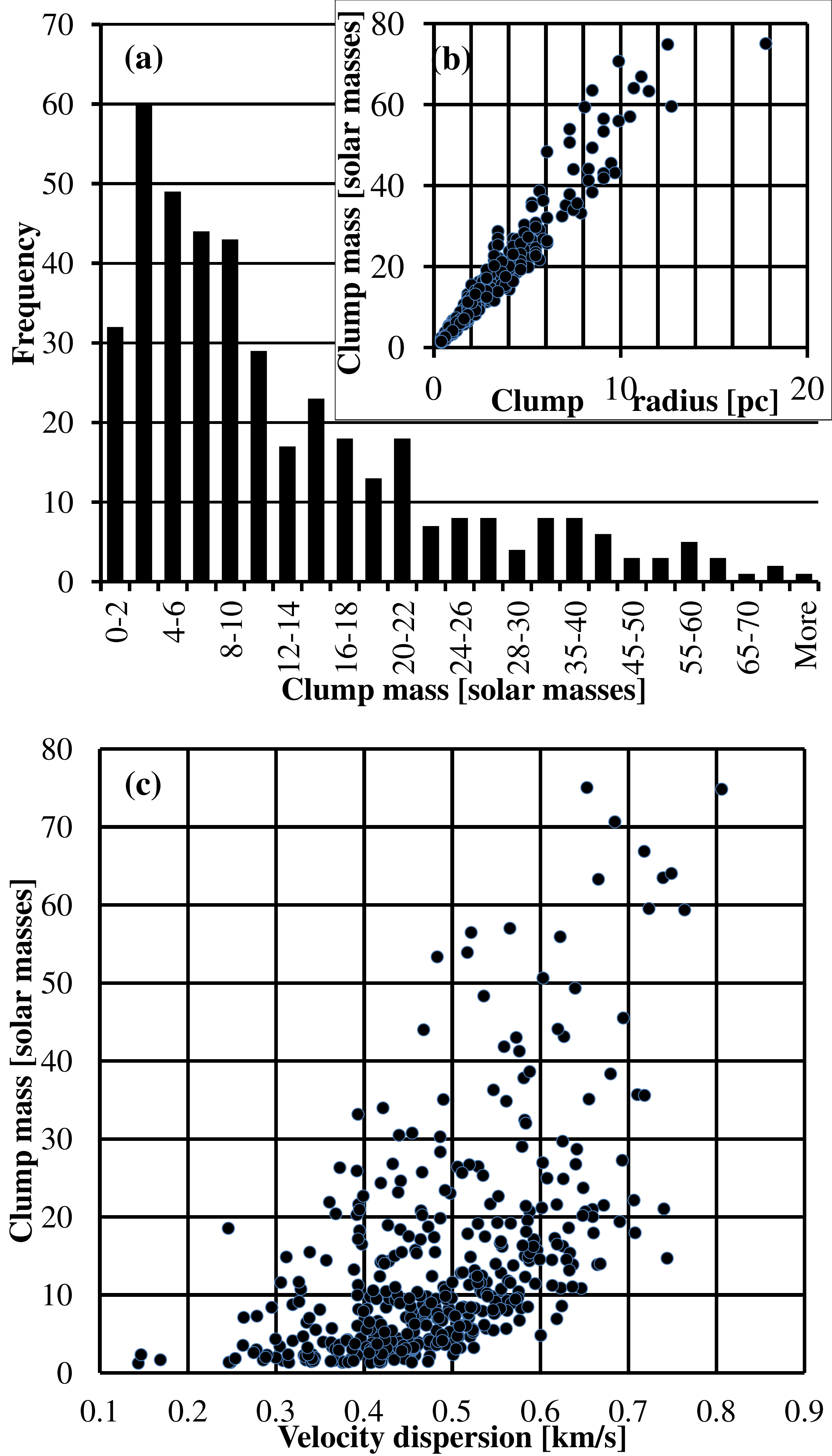}
\caption{Clump properties from the non-magnetic simulation. 
(a) mass distribution of 430 clumps. (b) clump mass vs.
clump radius. (c) clump mass vs. velocity dispersion inside the clump.} 
\label{fig8}
\end{figure}

We have used an algorithm developed for MG \citep*{vanloo15} in order to identify high density
clumps in the 3D $\beta=\infty$ field case with self-gravity. Setting a density threshold
of 100 H cm$^{-3}$, this algorithm scans through the computational domain and firstly
identifies all the cells that have a density above this threshold and then within that
selection identifies cells that neighbour each other to identify each clump as a whole. We have
discounted clumps containing only 1 grid cell from this data analysis. The algorithm
identified $\sim$430 clumps (not including a further 40 single cells with densities above the
threshold value). In Fig \ref{fig8}(a) we show the clump mass distribution of these $\sim$430 clumps.
Nearly half of the clumps have a mass less than 10\,M$_\odot$. The distribution has a 
defined peak in the 2-4\,M$_\odot$ bin and a long tail, typical of a log-normal 
distribution, with three clumps with masses greater than 70\,M$_\odot$.

In Fig \ref{fig8}(b) we show the distribution of clump mass vs. radius. There is a strong
agreement between clump mass $M_c$ and clump radius $R_c$ whereby 
$M_c\sim6\pm1.5\,R_c$ M$_\odot$. The clump radius ranges up to $\sim$18\,pc, 
although the next most massive clump has a radius of just over 12\,pc {\it and equivalent 
mass}; the three most massive clumps range widely in radius from 10 to
18\,pc. The average density across all the clumps is 4.9\,M$_\odot$\,pc$^{-3}$, with a 
standard deviation of 0.96\,M$_\odot$\,pc$^{-3}$, a maximum value of 8.35\,M$_\odot$\,pc$^{-3}$
and a minimum value of 3.05\,M$_\odot$\,pc$^{-3}$.

Turning now to the distribution of clump mass vs. velocity dispersion shown in Fig \ref{fig8}(c), 
a conclusion that can be reached is that lower mass clumps tend to have a lower velocity dispersion.
That said, there is a wide range of values even for the low mass clumps, from 0.1 to 
0.6 km\,s$^{-1}$. High mass clumps have velocity dispersions in the range 0.5 to 0.8
km\,s$^{-1}$. Compared to the observed distribution of \cite{heyer04}, 
the clumps measured here appear to have lower velocity dispersions for their sizes.
Velocity dispersions are underestimated by a factor of $2.8\pm0.67$ compared to the
scaling law of \cite{larson81} and $2.68\pm0.63$ compared to the scaling law of
\cite{solomon87}. We have examined the variation of these factors over time in our simulation and once
the clumps have stabilised (by t=0.055) these factors remain remarkably constant. Only
during the final collapse of the cloud (from t=0.085) into a single 17,000 M$_\odot$
clump/core does the velocity dispersion then fit with the Larson and Solomon et al. scaling 
laws, although taking a different threshold value will change sizes but not the velocity
dispersion and hence affect this comparison. 
In these non-magnetic simulations, under this examination it can be concluded that the
TI does not generate large enough velocity dispersions and so there is a need for
additional velocity dispersion from other sources. Further simulations will address whether 
this issue can be rectified with the TI alone, or whether feedback is required. The end-point
of our non-magnetic simulation would certainly suggest that the TI in combination with 
gravitational contraction can reproduce realistic velocity dispersions. A recent observational
test would suggest that gravity is the ultimate source of such velocity dispersions 
\citep{krumholz16}.

The gas temperatures in the clumps are on the order of 50-60K. Observationally inferred 
temperatures are lower than this. This would appear to highlight a limitation
of the cooling function and resolution used in this work. Higher resolution and
accounting for CO formation is likely to lead to higher densities and lower temperatures
but requires complex additions that are not within the scope of this work. We plan to study this in a future work.

As has been demonstrated earlier, a magnetic field suppresses the formation of clumps, 
so that filamentary structure tends to form instead. With equality of the magnetic pressure and thermal 
pressure ($\beta=1.0$), numbers of clumps and filaments form roughly equally. When the magnetic 
pressure dominates over the thermal pressure, no clumps form. In our 
simulations, with a low initial density and hence low initial pressure, this occurs even
for an average Galactic field strength. In these simulations, the magnetic field is as
yet unaffected by the formation of clumps. No field strength enhancement is observed.

It should be noted that under the definition in Table 1 of \cite{bergin07}, our ``clumps" 
are more like small Bergin \& Tafalla clouds in terms of their size and density, but more like 
Bergin \& Tafalla clumps in terms of their mass and velocity dispersion. We do not form 
structures in these simulations which can be considered as Bergin \& Tafalla cores.

\subsection{Filaments}

\begin{figure}
\centering
\includegraphics[width=80mm]{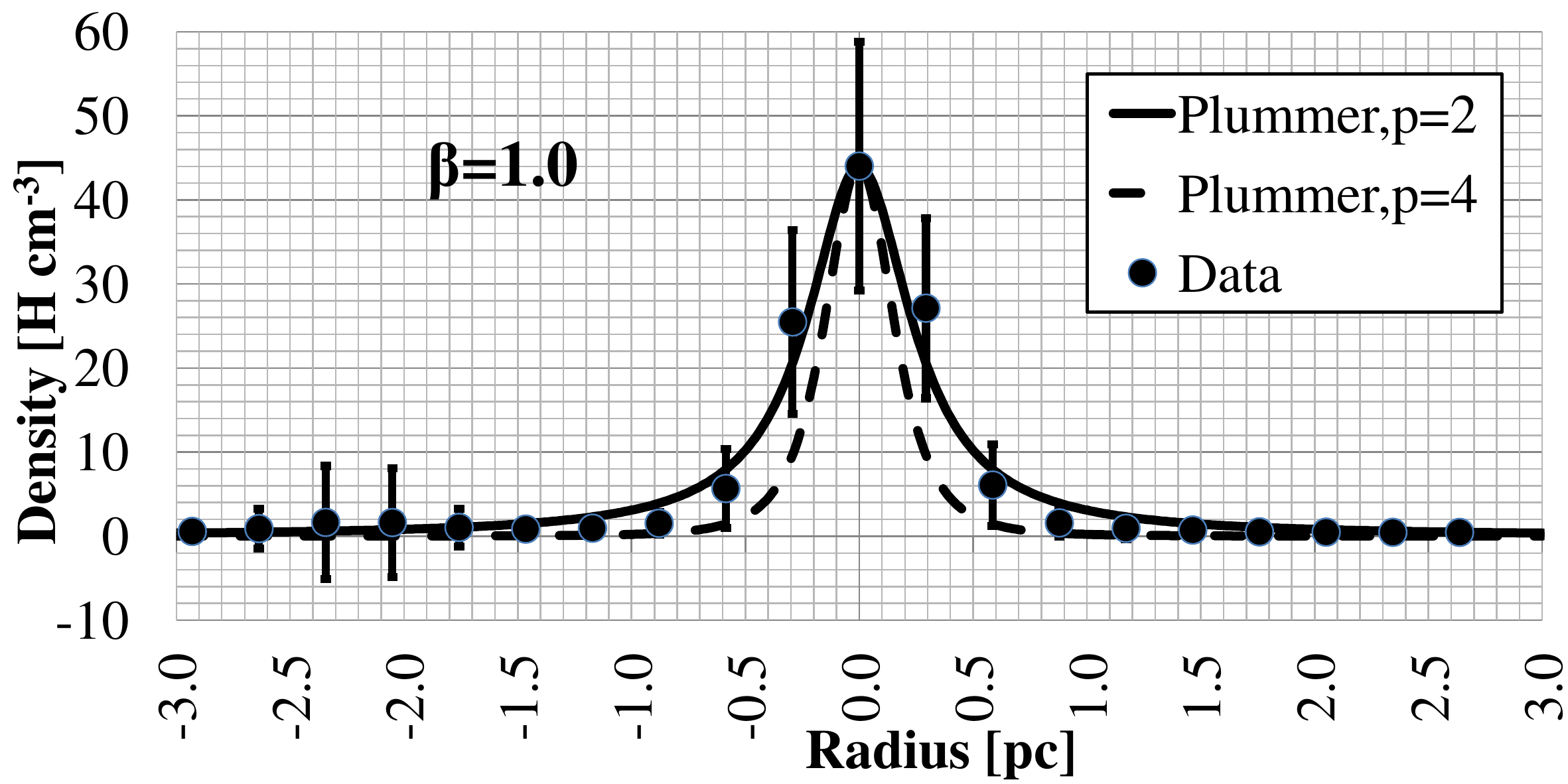}
\caption{Averaged filament profile across 36 filament cuts in the $\beta=1.0$ field 
case with self-gravity, compared to Plummer profiles (solid and dashed lines). 
The central peak density represents the average central
density across these filaments, with the error the standard deviation. The other data
points are generated from normalised filament profiles (normalising each filament
profile by its central density) in order to obtain these data points that are averages
and standard deviations across all 36 filament cuts.} 
\label{fig9}
\end{figure}

\begin{figure}
\centering
\includegraphics[width=80mm]{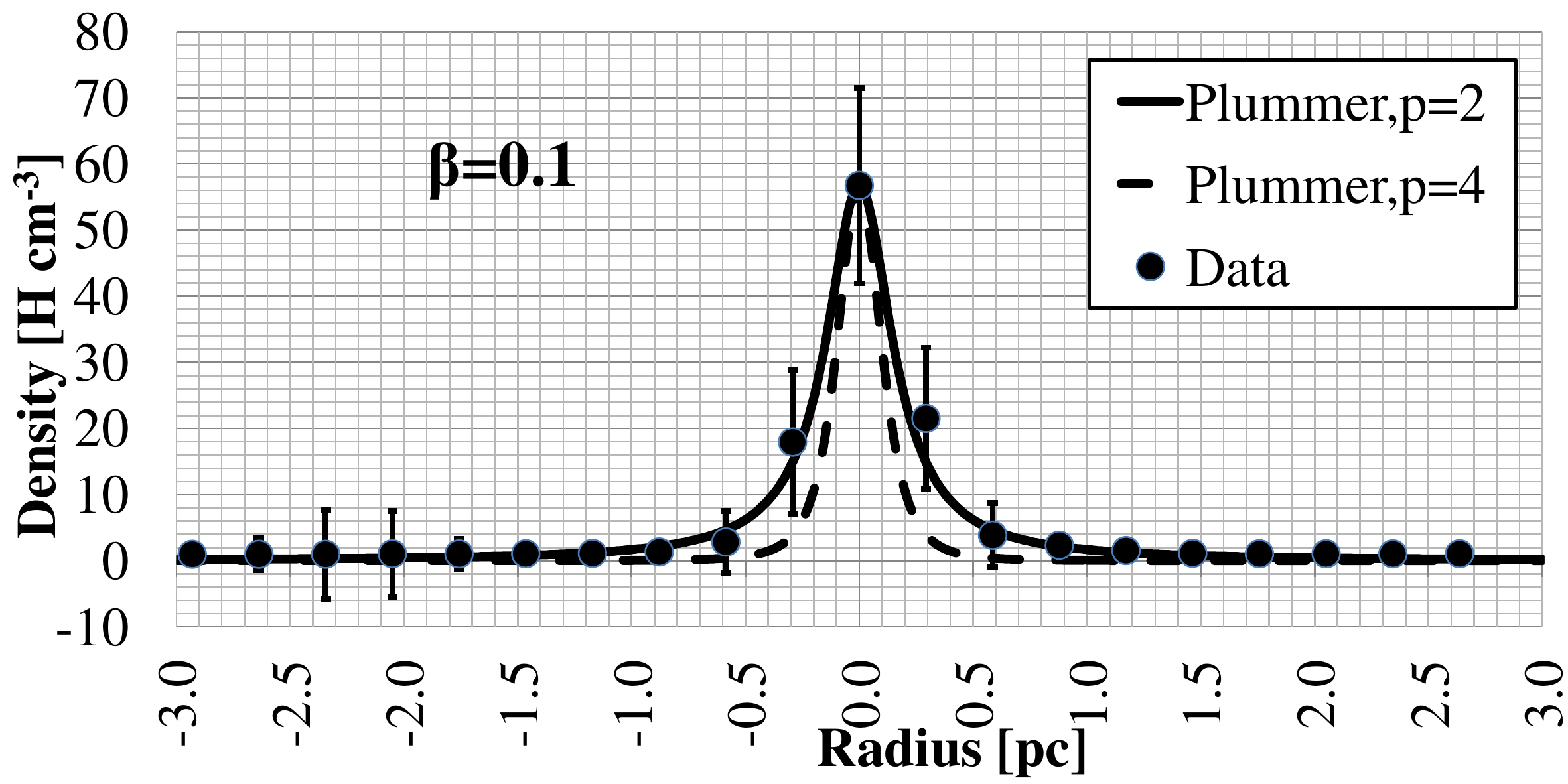}
\caption{Averaged filament profile across 26 filament cuts in the $\beta=0.1$ field 
case with self-gravity, compared to Plummer profiles (solid and dashed lines).
The data has been averaged and presented in the same way as Fig. \ref{fig9}.} 
\label{fig10}
\end{figure}

Filamentary structures can be described by a 
Plummer-like function of the form previously described with $p\approx2$ 
\citep{arzou11}. We now consider whether the filaments
formed in our simulations can be described by such a function and if so, 
with what $p$ value. In Figs \ref{fig9} and \ref{fig10} we show the averages of profiles
cut across filaments in the $\beta = 1.0$ and 0.1 field cases respectively.
These profiles are taken across the cloud at $y=0$ and values of $z$ ranging
across the $z$ range of the cloud, at regular intervals. The central density 
represents the average from the 36 filament measurements from the $\beta=1.0$ 
simulation and 26 filament measurements from the $\beta=0.1$ simulation. Fewer filament
profiles are identified and averaged over in the $\beta=0.1$ case as there are fewer
filaments in the cloud - see column 2 of Fig \ref{fig6}. For both values of $\beta$, the error on 
the central data point is the standard deviation of these central densities. The other data 
points are averages and the standard deviation calculated after each filament cut has been
normalised by its central density in order to allow such a data reduction across 
filaments with different central densities. In both field cases, we fit the data 
to Plummer profiles and show that a $p$=2 profile fits the data better than a $p$=4 
profile for $R_{flat}$ equal to 0.275\,pc in the $\beta=1.0$ case with FWHM
$\sim$0.6\,pc and $R_{flat}$ equal to 0.15\,pc in the $\beta=0.1$ case with FWHM
$\sim$0.35\,pc. These values of $R_{flat}$ and FWHM are consistent with 
other authors \citep{hennemann12,juvela12}.
For $\beta=1.0$, taking a filament with central density of 60\,cm$^{-3}$ and central 
temperature of 50\,K (and hence c$_s\sim$0.8\,km\,s$^{-1}$) from the profile at $y=0$, $z=0$, 
the Jeans length is $\lambda_J = C_s (\pi / G \rho)^{0.5} = 18.4$\,pc. For $\beta=0.1$, 
the same profile at $y=0$, $z=0$ has a filament with central density of 67\,cm$^{-3}$ and
central temperature of 50\,K, corresponding to a slightly smaller Jeans length $\lambda_J=17.5$\,pc.
Gravity is not yet affecting the shape and structure of these filaments. Formation is 
entirely due to the TI. With an enhanced cooling prescription the filaments will cool further 
and their density will increase. Eventually we expect gravity to dominate the evolution and 
the filaments will contract further on timescales shorter than the cloud evolution time.

The strength of magnetic field has an effect on filament properties. Comparing Figs \ref{fig9} and
\ref{fig10}, filaments formed in the strong magnetic field case are clearly narrower than in the
magnetic/thermal pressure equality case. This could be an effect of time-evolution, but
comparing the cloud evolution between $\beta=1.0$ and $\beta=0.1$ cases indicates this
is not the case. It is more likely due to the more-ordered evolution of the $\beta=0.1$ case,
forming a single corrugated sheet, as opposed to the ``bundle" of filaments in the 
$\beta=1.0$ case. As for the orientation of the
filaments, they are generally perpendicular to the magnetic field, which compares well
with observations. Filaments parallel to the field are only flows onto the denser 
perpendicular filaments in this scenario. In these simulations, the magnetic field is as
yet unaffected by the formation of filaments. No field strength enhancement is observed.
Further gravitational collapse of the filaments would presumably begin to affect the 
magnetic field on local filament scales, but larger field strengths measured across the cloud
must have a different origin in this scenario.

Temperatures in the filaments are similar to those of the clumps, and in some cases cooler:
at the highest densities, the temperatures approach 30\,K. Again, these are high compared
to the observations, for reasons given in the previous sub-section, but if these are
proto-filaments, the continued action of the TI under the influence of gravitational collapse
could lead to lower temperatures and higher densities. We hope to be able to examine this
in future work.

The velocity profiles across the filaments in both the $\beta=1.0$ and 0.1 simulations are complex.
Different filaments have different central velocities with non-thermal dispersions on
these central velocities. Typical filament velocities in the frame of reference of the entire cloud
are on the order of 5\,km\,s$^{-1}$ in both field strength cases, with velocities
generally directed towards the centre of the cloud. Typical velocity dispersions internally
within each filament are on the order of 0.5\,km\,s$^{-1}$. It is important to note three
things. Firstly, these internal velocity dispersions are on the order of those measured from
observations \citep{arzou11} - TI-driven filament formation can reproduce the ``turbulent"
velocities measured in filaments \citep[e.g. by][]{arzou13}. Secondly, the fact that different filaments have different
central velocities, whilst still being part of a larger bundle of filaments or even projection
of interconnected sheets highlights that it is now impossible to conclude from observations
of filaments with different central velocities that they are separate filaments - in this
work they could equally be part of a TI-formed corrugated sheet, non-uniform in density,
temperature and velocity. Thirdly, the entirely local TI mechanism drives large-scale, ordered flow on the same
scales as that observed and previously considered to be difficult to generate internally.
It should also be noted that at the width of the filaments ($\sim$0.3\,pc) in this 
study, a suitably-initialised turbulent velocity scaling relation would likewise predict velocity 
dispersions of $\sim$0.4\,km\,s$^{-1}$ so the thermal instability is not the only method to 
obtain such low velocity dispersions.

It should be noted that the clump/filament-formation timescale is far shorter than the diffuse medium evolution
timescale. The diffuse cloud initial condition in this work evolves on the scale of tens of Myrs, with a free-fall 
time t$_{ff}\sim$50\,Myrs. The medium remains quiescent and diffuse for the first $\sim$18\,Myrs,
after which time clumps and filaments characteristic of molecular clouds begin to form out of 
this medium, which given we start with the lowest possible density initial 
condition and introduce no external stimulus, is in good agreement with timescales from Galactic 
scale simulations \citep[section 2.6 of][]{ostriker10,kim11}. Hence, the giant molecular cloud can 
said to have formed at $\sim$18\,Myrs. The filaments then reach their final thermodynamic state 
with high density rapidly after this time (although slow movement of the filaments continues 
after this time) leading to star formation in the cloud on realistic timescales, in agreement
with observations and other numerical work \citep{clark12}. In our next work, we 
consider the injection of stars at this time and the effect of their wind and SN feedback onto the cloud, 
including timescales of cloud destruction. Given that massive stars will evolve to SN phase on 
timescales of 3-4\,Myrs, the cloud formed in these simulations will be strongly affected, if not
destroyed. We examine exactly this evolution and timescale in our next work.

\section{Conclusions}\label{conclusions}

In this paper we explore the idealised evolution of diffuse clouds under the influence of the thermal
instability \citep{field65}, with magnetic fields and self-gravity and without turbulence. We show that compared to the 
zero magnetic field case where symmetric stationary clumps rapidly form throughout the cloud, 
filaments extended perpendicular to the magnetic field lines form as material moves along 
the field lines. Over a longer time, the filaments continue to move, interconnect and disconnect 
in both 2D and 3D simulations. At any particular instant, projecting the 3D structure onto a plane 
generates column density projections that resemble filaments and bundles of filaments in 
molecular clouds if the plane is parallel to the magnetic field (i.e. \textbf{B} is
perpendicular to the line of sight). Projected column densities are comparable to observations.
Projecting the 3D structure onto a plane perpendicular
to the magnetic field (i.e. \textbf{B} is parallel to the line of sight), generates clouds remarkably
uniform in appearance, circular in projection and not at all filamentary, especially so in the strong
field case where material only moves along magnetic field lines and the result is the collapse of a
uniform spherical cloud {\it along this direction}. Column densities are 10$\times$ lower.

The action of the thermal instability in the formation of molecular clouds is not new, but the novelty
of this work is in the exploration of the operation of the thermal instability under the influence of the 
magnetic fields in numerical simulations of diffuse clouds, without external trigger
factors e.g. colliding flows or driven turbulence. Clumps and filaments can be formed under the action of
the thermal instability alone, without gravity, on realistic timescales with a range of properties comparable to those 
observed and without the need to resort to an initial ``turbulent" state. In fact, from stationary 
conditions, velocities such as those observed in molecular clouds are formed in these simulations.
This work further emphasizes the fact that the role of the thermal instability and magnetic fields 
should be fully considered in the formation of molecular clouds. The assumption of isothermal
conditions may be missing a significant mechanism. 

Limitations, such as neglecting the role of CO, can be interpreted as the reason for not matching 
all available observational results and future work including the thermal instability should strive to
address these limitations. We plan to study the particular role of CO in a future work.

Thermal instability leading to filament formation with or without the influence of gravity does not  
intensify the magnetic field in our simulations. Other influences are required, for example stellar
feedback generating super-sonic, super-Alfv{\'e}nic motions. This
particular question regarding the field-intensifying effect of stellar feedback will be explored 
in a forthcoming paper that takes these simulations as a starting point and introduces stars
and their wind and SNe feedback.

\section*{Acknowledgments}

This work was supported by the Science \& Technology Facilities Council
[Research Grant ST/L000628/1]. We thank the anonymous
referee for the positive review and minor suggestions which
have improved the manuscript.
The calculations for this paper were performed on the DiRAC Facility jointly 
funded by STFC, the Large Facilities Capital Fund of BIS and the University 
of Leeds. This facility is hosted and enabled through the ARC HPC resources 
and support team at the University of Leeds (A. Real, M.Dixon, M. Wallis, 
M. Callaghan \& J. Leng), to whom we extend our grateful thanks.
VisIt is supported by the Department of Energy with funding from the 
Advanced Simulation and Computing Program and the Scientific Discovery 
through Advanced Computing Program.

\label{lastpage}

\end{document}